\documentclass[]{interact}

\usepackage{epstopdf}
\usepackage[caption=false]{subfig}

\usepackage{amssymb}
\usepackage{amsthm}
\usepackage[ruled,vlined]{algorithm2e}
\usepackage{breqn}
\usepackage{cellspace}
\usepackage{hyperref}
\usepackage{bbm}
\usepackage{graphicx}
\usepackage{float}
\graphicspath{ {./Figure/} }
\usepackage{tikz}
\usetikzlibrary{calc,3d}
\usepackage{diagbox}
\usepackage{thmtools, thm-restate} 
\usepackage[utf8]{inputenc}
\usepackage[english]{babel}
\usepackage{comment}
\usepackage{cleveref}

\def\R{{\mathbb{R}}}   
\def\Z{{\mathbb{Z}}}   

\def\N{{\mathbb{N}}}
\def\E{{\mathbb{E}}}
\def\P{{\mathbb P}}
\def\1{{\mathbf{1}}}

\def\A{{\mathcal A}}  
\def\H{{\mathcal H}}   
\def\Q{{\mathcal Q}}   
\def\K{{\mathcal K}}   
\def\M{{\mathcal M}}   
\def\F{{\mathcal D}}   

\declaretheorem[name=Assumption,
refname={assumption,assumptions},
Refname={Assumption,Assumptions}]{defassump}
\declaretheorem[name=Theorem,
refname={theorem,theorems},
Refname={Theorem,Theorems}]{defthm}
\declaretheorem[name=Lemma,
refname={lemma,lemmas},
Refname={Lemma,Lemmas}]{deflemma}
\declaretheorem[name=Definition,
refname={definition,definitions},
Refname={Definition,Definitions}]{defdef}
\declaretheorem[name=Remark,
refname={remark,remarks},
Refname={Remark,Remarks}]{defrmk}
\declaretheorem[name=Corollary,
refname={corollary,corollarys},
Refname={Corollary,Corollarys}]{defcoro}

\usepackage[natbibapa,nodoi]{apacite}
\begin{document}

\articletype{Research Article}

\title{Optimal Acceptance of Incompatible Kidneys}


\author{
\name{Xingyu Ren\textsuperscript{a}, Michael C. Fu\textsuperscript{b}, and Steven I. Marcus\textsuperscript{a}\thanks{CONTACT Xingyu Ren. Email: renxy@umd.edu}}
\affil{\textsuperscript{a}Department of Electrical and Computer Engineering \& Institute for System Research, University of Maryland, College Park, USA\textsuperscript{b}Robert H. Smith School of Business \& Institute for System Research, University of Maryland, College Park, USA}
}

\maketitle

\begin{abstract}
    Incompatibility between patient and donor is a major barrier in kidney transplantation (KT). The increasing shortage of kidney donors has driven the development of desensitization techniques to overcome this immunological challenge. Compared with compatible KT, patients undergoing incompatible KTs are more likely to experience rejection, infection, malignancy, and graft loss. We study the optimal acceptance of possibly incompatible kidneys for individual end-stage kidney disease patients. To capture the effects of incompatibility, we propose a Markov Decision Process (MDP) model that explicitly includes compatibility as a state variable. The resulting higher-dimensional model makes it more challenging to analyze, but under suitable conditions, we derive structural properties including control limit-type optimal policies that are easy to compute and implement. Numerical examples illustrate the behavior of the optimal policy under different mismatch levels and highlight the importance of explicitly incorporating the incompatibility level into the acceptance decision when desensitization therapy is an option.
\end{abstract}


\section{Introduction}\label{sec1}

Kidney transplantation is the organ transplantation of a donated kidney into a patient with end-stage kidney disease (ESKD). For most ESKD patients, transplantation is their best option. Compared with those undertaking dialysis treatment in their remaining lifetime, patients who undergo transplantation usually live longer and have better quality of life \citep{aimaretti2016preemptive}. To receive cadaveric kidney offers, ESKD patients in the U.S. have to join the waitlist of the United Network for Organ Sharing (UNOS), which manages the Organ Procurement and Transplantation Network (OPTN). The OPTN kidney allocation system (KAS) works as follows \citep{optnpolicy2023}. Once a kidney is available, the UNOS will identify the matched candidates on the waitlist and offer the kidney to the patient with the highest priority, which is determined by the medical and listing status of both patients and donors (e.g., patient waiting time and the distance between the patient and donor). The transplant surgeon responsible for the care of the patient has a very short time to make a final decision on whether or not to accept the kidney for transplantation. If the kidney is declined, it will be offered to another eligible patient in the descending order of priority. Patients declining an offer will maintain their status on the waitlist without being penalized, and may even gain a higher priority on the waitlist in the future due to increased waiting time. Although the number of kidney donors has been increasing, it is eclipsed by the number of newly-listed transplantation candidates. According to the OPTN database \citep{optn2020}, there were 27332 candidates receiving kidney transplantation during 2023, while 44565 candidates were added to the waitlist. By the end of 2023, there were $88667$ candidates on the waitlist. Once added to the waiting list, candidates usually have a lengthy wait, with an average waiting time of $2.13$ years since the new OPTN allocation policy was implemented in $2020$ \citep{unos2022}. Fewer than half of candidates eventually receive transplantation, and more than $5000$ candidates die on the waitlist every year.

To improve access to kidney transplantation, modern desensitization techniques have been developed to overcome the human leukocyte antigen (HLA) and the ABO blood-type incompatibility (HLAi and ABOi), the major immunological barriers to kidney transplantation \citep{konvalinka2015utility,rydberg2007abo}. The HLA antigens are polymorphic proteins, and the ABO antigens consist of oligosaccharides expressed on donor kidney allograft. They are potential targets for the immune system of the organ recipients. 
The extent of sensitization to the HLA antigens is reflected by the calculated panel reactive antibody (CPRA) value of a patient, which is the proportion of donors \textit{expected} to have HLA mismatch with that patient. Patients with CPRA value greater than 0 are called sensitized. About $40\%$ of patients on the waitlist are sensitized \citep{unos2022}. Compared with insensitive patients, highly sensitized patients have a much lower chance of finding a compatible donor: it may take years without a compatible donor being identified \citep{kuppachi2020desensitization}. 

To receive an incompatible kidney, a patient has to undergo desensitization therapies that aim to reduce or remove donor specific antibodies (DSA) prior to and after the transplantation. With the development of modern desensitization protocols, satisfactory outcomes have been observed in both HLAi and ABOi kidney transplantations. However, HLAi and ABOi kidney recipients show lower graft and patient survival rates compared to recipients of compatible kidneys \citep{optn2020, koo2015current, kim2021does, morath2017abo}. Incompatible kidney transplantation comes with distinct drawbacks, as desensitization therapies increase the risk of infection and malignancy \citep{clayton2017sensitized}. Moreover, recipients of incompatible kidneys are more prone to experiencing acute or chronic rejection and early graft loss, in contrast to recipients with compatible kidneys \citep{koo2015current, ko2017clinical}. Even so, incompatible kidney transplantation still remains the best therapeutic option for patients who are highly sensitized and/or difficult to match, and in recent years, HLAi transplantation has become more common. According to the OPTN data from $2017$ to $2021$ \citep{srtrkidney2021}, more than $70$\% of deceased donor kidney recipients, and more than $40$\% of living donor kidney recipients had four or more HLA mismatches. On the other hand, ABOi kidney transplantation is less common: the current OPTN kidney allocation policy allows deceased donor ABOi transplantation only under certain circumstances, e.g., when the HLA mismatch level is zero \citep{optnpolicy2023}.

For patients with an incompatible directed living donor, they have the option of joining the Kidney Paired Donation (KPD) program to find a compatible living donor. The OPTN tracks every incompatible donor-recipient pair that registers in the KPD program, and the UNOS works with transplant centers to identify all possible matches where the donor in each pair is compatible with the recipient in another pair \citep{ashlagi2021kidney}. By a chain of exchanging donors through multiple pairs, a compatible match for all the recipients can be established. KPD is an effective means for offering better matched organs to patients and thus reducing the kidney shortage, and the proportion of paired donations in living donor kidney transplantations has grown from $12.0\%$ to $18.6\%$ in the past five years. Furthermore, incompatible transplantation could be incorporated into to the KPD program to further augment its benefits. For example, currently, ABOi transplantation is not allowed in the KPD program \citep{optnpolicy2023}, whereas by incorporating incompatible transplantation into the KPD program, a more flexible and adaptable matching/pairing system can be established, potentially leading to a higher number of successful transplantations.




Although the kidney shortage is severe, it is reported that most cadaveric donor kidney offers are declined by at least one transplant surgeon before being accepted for transplantation \citep{husain2019association}, and $25\%$ of offers are eventually discarded by transplant surgeons \citep{unos2022}. Unsatisfactory organ quality accounts for most of the declined offers. Meanwhile, the number of kidney transplant recipients who experience graft failure and return on dialysis has been increasing every year \citep{fiorentino2021management}. To improve systemwide health outcomes, researchers have taken the perspective of a policy maker and modeled the organ allocation problem as a multi-class queueing problem or a sequential allocation problem \citep{tuncc2022simple,ata2017organjet,akan2012broader,su2004patient,su2005patient,su2006recipient}, focusing on the overall social welfare (e.g., the total organ usage) and the trade-off between efficiency and equity of the allocation system; as a result, simplified models are used for acceptance decisions at the patient level, for example, the multi-class queueing models of \citet{tuncc2022simple,akan2012broader} and \citet{su2005patient,su2006recipient} categorize patients into different types based on their medical conditions and listing statuses, but assume their types unchanged over time and/or represent their acceptance strategy by the probability of acceptance. While this static classification or simplified strategy might be adequate for the purpose of kidney allocation, from the perspective of an individual patient, whether or not to accept a kidney offer is a critically important decision affecting their quality of life, motivating our focus on the decision-making process at the patient level.

Thus, our research models the transplant surgeon (the decision maker) decision-making process to capture the dynamic nature of the patient state, ensuring that the decision to accept a possibly incompatible kidney offer aligns with the patient's specific circumstances. When a kidney offer arrives, the decision maker has to decide whether to accept it, depending on the current listing and medical state of the patient and the characteristics of the kidney offer, including both quality and compatibility. If the kidney is accepted, the patient will undergo transplantation; otherwise, the patient will wait for the next offered kidney and the current offer is no longer available. Our research primarily focuses on patients on the cadaveric organ waitlist, but the same decision-making process extends to individual donor-patient pairs in the KPD program. Therefore, our research also applies to modeling and analyzing the decision-making behavior of individual pairs in the KPD program.

Current support tools for kidney transplant decision making, e.g., logistic regression models available at the website of the Scientific Registry of Transplant Recipients (SRTR) \citep{srtrAid}, and the Transplant Models website developed by the Center for Surgical \& Transplant Applied Research (C-STAR) at NYU Langone \citep{bae2019cansurvival,bae2019can,grams2013,grams2012candidacy}, only consider current characteristics of a fixed donor-recipient pair and predict outcomes of the transplant surgery, without including various uncertainties that the decision maker faces, including future patient state and future availability of kidney offers. To model the basic trade-off between waiting for a higher-quality and/or more compatible kidney and the risk of deterioration of health while waiting, 
we propose a Markov decision process (MDP) model to study this problem \citep{bertsekas2020dynamic}.


Although similar problems have been studied in the context of liver transplantation \citep{alagoz2004optimal,alagoz2007determining,alagoz2007choosing,alagoz2010markov,kaufman2017living,batun2018optimal}, there are several major differences between liver and kidney transplantation. First, dialysis is an alternative option for ESKD patients, so the urgency may not be as severe (hence the patient remaining lifetime can be measured in months or years rather than in days), while liver transplantation is the only available therapy for end-stage liver disease (ESLD) patients. Moreover, HLA incompatibility is a major barrier in kidney transplantation, but the effect of HLA incompatibility is unclear in liver transplantation \citep{mahawar2004role}. ESLD patients under urgent medical condition (i.e., those having high model for end-stage liver disease (MELD) scores) are prioritized in the liver allocation system, while the patient sensitivity level and waiting time also play an important role in kidney allocation.

In the setting of kidney transplantation, \citet{david1985,ahn1996involving,bendersky2016deciding,ren2023sensitivity} propose MDP models for the optimal acceptance of kidneys for individual ESKD patients, and a more recent paper \citep{fan2020optimal} studies the optimal timing to start dialysis treatment and accept a kidney offer (see \cite{ren2022review} for a more comprehensive review of MDP models on individual patient organ acceptance decision making). Although organ quality and compatibility are key factors in kidney transplantation \citep{koo2015current,bae2019can}, previous work has modeled them in an {\it implicit} manner and considered only one or the other of the two factors -- not both simultaneously -- which can lead to decisions that are suboptimal. For example, an MDP model that takes only the compatibility into consideration is likely to reject an offer of low compatibility but high quality, which could have been the best choice for the patient, especially with the option of current desensitization therapies. Moreover, previous research has summarized all the short-term and long-term effects of the transplantation in a terminal reward, i.e., treating transplantation as a terminal state. As mentioned previously, low-quality or incompatible kidney recipients are more likely to encounter transplantation failure due to infection, rejection, and early graft loss, which would necessitate a return to dialysis and a desire for retransplantation shortly thereafter (e.g., within several months). Therefore, an MDP model explicitly modeling the retransplantation is desirable.

We propose an MDP model that incorporates the option of incompatible kidney transplantation via desensitization therapies. By including both the quality and the mismatch level {\it explicitly} as state variables in the kidney acceptance decision process, our model captures cost-benefit trade-offs between the quality and the compatibility of the kidney offer, and between waiting for a compatible kidney versus receiving an earlier transplantation but having to undergo desensitization treatment. We also explicitly model a patient who returns to being on dialysis and rejoins the waitlist for a retransplantation after experiencing an early graft loss, whereas previous work simply terminates the decision process upon organ acceptance. In particular, we model the probability of a transplantation failure to be a function of the patient state and both quality and compatibility of the donor kidney. Consequently, the state vector has a more complex correlation structure and the state dynamics are more complicated, which makes it more challenging to characterize the form of optimal policies and prove structural results.

In summary, our model takes into consideration various uncertainties and trade-offs, such that the long-term benefit of accepting or rejecting an incompatible kidney offer can be quantified, which not only provides transplant surgeons with a decision-support tool, but also promotes the utilization of incompatible kidney transplantation, which may further expand the donor pool. Moreover, our research results can be integrated into the design of both KAS and KPD policies \citep{tuncc2022simple,ata2017organjet,su2004patient,su2005patient,su2006recipient}, where the individual patient decision-making procedure is an important component, e.g., in \citet{tuncc2022simple,su2004patient}, individual patients are assumed to adopt a control limit-type policy.

To summarize, the main contributions of this paper include the following:
\begin{enumerate}
    \item In terms of modeling, our model is the first to incorporate both the quality and the compatibility of the kidney offer and to explicitly model transplantation failure and retransplantation. As a result, we are able to quantify the tradeoffs between the opportunity to get off dialysis earlier but having to undergo desensitization treatment with a higher possibility of a poor transplant outcome. 
    \item In terms of theory, we are able to prove desirable structural properties of the resulting more complicated MDP model under realistic conditions. In particular, we establish sufficient conditions for the existence of control limit-type optimal policies and identify situations where a control limit-type optimal policy does not exist if some condition is violated.
    \item In terms of practice, preliminary numerical experiments that quantify the improved health outcomes illustrate the impact of incorporating both the quality and the compatibility of the kidney offer and allowing the option of incompatible kidney transplantation with desensitization treatment. The results indicate improvement on the order of an additional year of life expectancy for elder ESKD patients, representing a substantial gain, since
    their expected remaining lifetime is less than five years without kidney transplantation.
\end{enumerate}

The rest of the paper is organized as follows: In \Cref{sec2}, we formulate the individual patient kidney acceptance problem as an MDP model. In \Cref{sec3}, under some intuitive assumptions, we derive structural properties including control limit-type optimal policies. In \Cref{sec4}, we conduct numerical experiments to evaluate the behavior of the optimal policy under different quality and mismatch levels and illustrate the impact of incorporating compatibility and retransplantation. \Cref{sec5} concludes the paper and points to future research directions. Proofs and details of parameter selection in the numerical experiments are included in the Appendix.

\section{Model Formulation}\label{sec2}
We formulate the individual patient kidney acceptance problem as a discrete-time, infinite-horizon MDP. The set of decision epochs is the natural numbers $\N=\{0,1,2,\cdots\}$, where the unit could be months (e.g., $1$ month or $6$ months). At each epoch $n\in \N$, the patient state is updated and at most one kidney offer may arrive. The decision to be made is whether to accept the offer based on the current patient state and both quality and compatibility of the offer. If the decision maker accepts the kidney and the transplantation is a success, the decision process terminates; otherwise, the patient waits for the next offered kidney. The decision process terminates when a successful transplantation happens or the patient dies. The objective is to maximize the total reward accumulated over the entire decision process.

\begin{defrmk}
    Patients typically receive their first kidney offer in about 80 days, and kidneys continue to become available approximately once a month \citep{husain2019association}. Since half of these offers don't meet the transplant center's criteria \citep{king2022role}, a decision period of one to six months is reasonable, though this varies by patient characteristics. For example, unsensitized patients receive twenty times more offers than highly sensitized ones, and blood type O patients receive five times more offers than those with blood type B (see Tables A.77 and A.78 in \citet{unos2022}). Thus, decision periods should consider factors like sensitivity level, blood type, and other patient characteristics.
\end{defrmk}

\subsection{State of Patient and Kidney Offer Stochastic Processes}
The state space is $\mathcal S:= S_H\times S_K \times S_M \bigcup \{P\}$. At each epoch $n$, the state $s_n$ is either a triple $(h_n,k_n,m_n)$, or the post-transplantation state $P$. 
\begin{itemize}
    \item $\{h_n\}$: \textbf{Patient state}. $h_n$ is a scalar summarizing patient listing and medical status and taking values in a finite set of positive integers $S_H=\{1,\cdots,H,H+1\}$, where a larger value implies worse patient state and $H+1$ represents death. For example, we can represent the patient state by the estimated post-transplant survival (EPTS) score \citep{bae2019can,unosEPTS}, which incorporates the patient's age, diabetes status, time on dialysis, etc., or other comprehensive indices.
    \item $\{k_n\}$: \textbf{Kidney offer state}. $k_n$ represents the quality (e.g., kidney donor profile index (KDPI) score for deceased donors \citep{unosKDPI} and Live Donor KDPI (LKDPI) score for living donors \citep{lkdpi2015,massie2016risk}, which incorporates the donor's age, height, weight, diabetes status, serum creatinine, etc.) of the donor kidney available at the current decision epoch and takes values in a finite set of positive integers $S_K=\{1,\cdots,K,K+1\}$, where a larger value implies worse quality and $K+1$ means that \textit{no} kidney offer is available.
    \item $\{m_n\}$: \textbf{Mismatch level}. $m_n$ measures the compatibility between the patient and the kidney donor (e.g., ABO and HLA mismatch level), and takes values in a finite set $S_M=\{1,\cdots,M\}$, where a larger value implies higher mismatch level and $1$ means perfect match.
    \item $P$: \textbf{Post-transplantation state}. The MDP will transition into the absorbing state $P$ if the patient undergoes a \textit{successful} transplantation. Without loss of generality, we may take $P\in \Z^3 \setminus \{S_H\times S_K \times S_M\}$, e.g., $P=(0,0,0)$ so that $\mathcal S\subset \Z^3$.
\end{itemize}

\begin{defrmk}\label{rmk:waiting}
According to the current OPTN kidney allocation policy \citep{optnpolicy2023}, an important factor for allocating deceased-donor kidneys is the patient waiting time. Though the waiting time is not explicitly modeled as a state variable, we can incorporate it into the patient state. For example, the formula for computing the EPTS score includes the time on dialysis, which often coincides with the patient waiting time \citep{unosEPTS,optnpolicy2023}.
\end{defrmk}

\begin{defrmk}
The kidney offer state $k_n$ only describes the \textit{donor} status, while the mismatch level $m_n$ is the compatibility of the donor-recipient pair. The patient needs to undergo desensitization treatment to accept an incompatible offer.
\end{defrmk}

\subsection{Action \& Action Space}
Let $\{a_n\}_{n\in\N}$ be the decision maker's \textit{actions}. For each $n\in\N$, $a_n\in \mathcal A=\{W,T\}$ where
\begin{itemize}
    \item $W$: reject the current offer and \textbf{wait} for one more period.
    \item $T$: accept the current offer for \textbf{transplantation}.
\end{itemize}
For any state $(h,k,m)\in S_H\times S_K \times S_M$, the feasible action set is
\begin{align*}
    A(h,k,m):=\begin{cases} \{W,T\} &k\neq K+1, \\ 
    \{W\} &k=K+1, \end{cases}
\end{align*}
i.e., when a kidney offer is available, the decision maker can either reject by choosing $W$, or accept by choosing $T$; the only choice is to wait if the kidney offer is unavailable.
\subsection{Dynamics}
\begin{itemize}
    \item $\H(j|i):=\P(h_{n+1}=j |  h_n=i , a_n=W), ~\H(\cdot|\cdot):S_H\times S_H \mapsto [0,1],$ is the probability that the patient state is $j$ at epoch $n+1$, given that the patient is in state $i$ and the decision maker chooses $W$ at epoch $n$. We set $\H(h|H+1)=0,~h=1,\cdots,H$ and $\H(H+1|H+1)=1$, i.e., ``death states" $\{(H+1,k,m),~\forall k ,m\}$ are absorbing.
    \item $\K(k_n|h_n),~ \K(\cdot|\cdot):  S_K\times S_H\mapsto [0,1],$ is the probability that a kidney of state $k_n$ is offered to a patient in state $h_n$. We assume that the distribution of the kidney state is only a function of the current patient state. We set $\K(K+1|H+1)=1$, i.e., the patient doesn't receive any offer after death.
    \item $\M(m_n),~\M(\cdot):S_M \mapsto [0,1],$ is the probability that the mismatch level between the patient and donor is $m_n$. The distribution of the mismatch level predominantly depends on the patient HLA characteristics and ABO blood type. By categorizing patients according to their HLA characteristics and blood type, we can assume that $\{m_n\}_{n\in\N}$ is an independent and identically distributed (i.i.d.) sequence of random variables, also independent of processes $\{h_n\}_{n\in\N}$ and $\{k_n\}_{n\in\N}$.
    \item $\F(h_n,k_n,m_n),~\F(\cdot,\cdot,\cdot): S_H\times S_K\times S_M \mapsto [0,1),$ is the probability of the transplantation failure for a patient in state $h_n$ transplanted with a kidney of state $k_n$ and mismatch level $m_n$. When the decision maker chooses to transplant, there are two possible outcomes: the transplantation is a success or a failure (e.g., early graft loss). With probability (w.p.) $1-\F(h_n,k_n,m_n)$, the transplantation is a success, the state transitions to the absorbing state $P$ and the decision process terminates. Otherwise, the patient returns to the waitlist for a retransplantation. For the latter case, the patient state is more likely to become worse, and will evolve according to the following transition law.
    \item $\Q(j|i):=\P(h_{n+1}=j |  h_n=i,~\text{transplant fails at }n),~\Q(\cdot|\cdot): S_H\times S_H \mapsto [0,1],$ is the probability that the patient state is $j$ at time $n+1$, given that the patient is in state $i$ and a transplantation failure happens at time $n$. We take the function $\Q$ as different from the function $\H$, considering that desensitization treatment and the transplantation failure may have a negative impact on the patient health.
\end{itemize}

\begin{defrmk}
Patient state $h_n$ is a comprehensive index computed by the medical and listing status of the patient, which also determine the priority ranking of the patient on the waitlist. For example, as mentioned in \Cref{rmk:waiting}, we may define the patient state by the EPTS score, which is strongly correlated with the patient waiting time, an essential factor of the priority ranking. Assuming the distribution of the kidney state to be a function of the patient state implies that the chance to get a kidney offer depends on the priority ranking on the waitlist, which implicitly models the effect of the kidney allocation system.
\end{defrmk}

\begin{defrmk}
To justify the assumption that $\{m_n\}_{n\in\N}$ is independent of $\{h_n\}_{n\in\N}$ and $\{k_n\}_{n\in\N}$, for illustration, we take the HLA mismatch level (as the current U.S. kidney allocation system matches only ABO blood-type compatible pairs), which is based solely on genetic differences between the donor and recipient. Since a patient’s HLA antigen type is fixed and unrelated to their health, it is reasonable to assume that the mismatch level is independent of the patient state transition.

When a donor organ becomes available in the allocation system, for a specific patient, the distribution of mismatch level between the patient and donor is determined solely by the population's HLA antigen distribution, independent of patient state or organ quality. The allocation score, which considers factors like organ quality, patient state, and mismatch level, prioritizes patients with the highest scores. However, the score only considers HLA mismatch levels of zero or one, ignoring mismatch levels of two or higher \citep{optnpolicy2023}. With $95\%$ of U.S. transplantations having a mismatch level of two or above \citep{unos2022}, the mismatch level has a negligible effect on both the likelihood of receiving a kidney offer and the quality of the kidney offer in most cases. Thus, assuming that ``compatibility is independent of the patient state and organ quality'' is a reasonable approximation.
\end{defrmk}

To summarize, we provide the overall decision making procedure in \Cref{flow}, and the general transition probability as follows: for any $n\in \N$,
\begin{align*}
    &~~~~\P(s_{n+1}=P  |  s_n=s,a_n=a)\\
    &=\begin{cases}
    1 &\text{if } s=P,\\
    1-\F(h,k,m) &\text{if } s=(h,k,m), a=T,\\
    0 &\text{otherwise};
    \end{cases}\\
   &~~~~\P(s_{n+1}=(h',k',m')  |  s_n=s,a_n=a)\\
   &=\begin{cases}
    \M(m')\K(k'|h')\H(h'|h) &\text{if } s=(h,k,m), a=W,\\
    \F(h,k,m)\M(m')\K(k'|h')\Q(h'|h) &\text{if } s=(h,k,m), a=T,\\
    0 &\text{otherwise}.
    \end{cases}
\end{align*}

\begin{algorithm}[bt]
\caption{Kidney transplantation decision making procedure.\label{flow}}
Initial patient heath $h_1$\;
\For{decision epoch $n=1,2,\cdots$}{
  Receive a kidney offer $(k_n,m_n)$ w.p. $\K(k_n|h_n)\M(m_n)$ \;
  Make a decision whether to accept the offer\;
  \If{\text{the offer is accepted}}{
    \eIf{\text{transplantation is successful}}{
    Receive terminal reward $r(h_n,k_n,m_n)$\; The process \textbf{terminates}\;
    }
    {
    The transplantation fails\;
    Receive reward $c(h_n)$\;
    The patient state transitions to $h_{n+1}$ w.p. $\Q({ h_{n+1}|h_n})$\;
    }
  }
  \If{\text{the offer is declined}}{
    Receive reward $c(h_n)$\;
    The patient state transitions to $h_{n+1}$ w.p. $\H({h_{n+1}|h_n})$\;
    }
   \If{\text{the patient dies, i.e. $h_{n+1}=H+1$}}{
    The process \textbf{terminates}\;
    }
}
\end{algorithm}
\subsection{Reward Functions}
\begin{itemize}
    \item $c(\cdot): S_H \mapsto \R_+$, \textbf{the intermediate reward function.} If a patient in state $h_n$ does not undergo a successful transplantation, i.e., the decision maker either chooses $W$ or chooses $T$ but the transplantation fails, they get an intermediate reward $c(h_n)$ for being alive for one period. We set $c(H+1)=0$.
    \item $r(\cdot,\cdot,\cdot): S_H\times S_K \times S_M \mapsto \R_+$, \textbf{the terminal reward function.} If a patient in state $h_n$ undergoes a successful transplantation with a kidney of quality $k_n$ and mismatch level $m_n$, they receive terminal reward $r(h_n,k_n,m_n)$. Reward function $r$ measures the long-term effect of a successful transplantation which terminates the decision process. We set $r(H+1,k,m)=0,~\forall k\in S_K,~m\in S_M$.
\end{itemize}

We assume that the reward accruing in the post-transplantation state $P$ is zero. For any feasible state-action pair $(s,a)$, if $a=T$, the one-stage reward $g(s,a)$ is given by
\begin{align*}
    g((h,k,m),T)=\begin{cases} r(h,k,m) & \text{w.p. } (1-\F(h,k,m)), \\ c(h) &\text{w.p. } \F(h,k,m). \end{cases}
\end{align*}
If $a=W$, the one-stage reward $g(s,W)=c(h)$ is a constant.

\begin{defrmk}
The death of the patient also terminates the decision process, since death states are absorbing and the patient receives zero reward upon death.
\end{defrmk}

\subsection{Objective Function}
The goal is to find a policy $\pi: S\mapsto \mathcal A$ that maximizes the expected total discounted reward
\begin{align*}
    f_\pi(h,k,m) :=\E\left(\sum_{i=0}^\infty \lambda^i g(s_i,\pi(s_i)) | s_0=(h,k,m) \right)
\end{align*}
for any initial state $s_0=(h,k,k)\in S_H \times S_K\times S_M$, where $\lambda\in[0,1]$ is the discount factor. We only consider stationary policies (i.e., policies that don't depend explicitly on time) in this paper. Denote the maximum expected total discounted reward (also known as the value function) by $V(h,k,m):=\max_{\pi \in \Pi} f_\pi(h,k,m)~,\forall h,k,$ and $m$, where $\Pi$ is the set of stationary policies.

\section{Structural Results}\label{sec3}
In this section, we present two types of structural results: monotonicity of the value function $V(h,k,m)$, and the existence of control limit-type optimal policies, which will be formally defined. First, we present \Cref{optimality} on the Bellman equation and convergence of the value iteration algorithm, which are standard MDP results and useful for computing the optimal policy and proving other structural results. As the MDP model has finite state and action spaces and is time-homogeneous, \Cref{optimality} follows from Proposition 5.4.1 in \citet{bertsekas2020dynamic}.
\begin{defthm}\label{optimality}
The value function $V$ satisfies: $\forall h\in S_H,~m \in S_M$, and $k< K+1$,
\begin{align} \label{bellman-dis}
\begin{split}
    &~~~~V(h,k,m)\\
    &=\max\begin{pmatrix}
    (1-\F(h,k,m))r(h,k,m)+\F(h,k,m) (c(h)+\lambda \sum_{h'\in S_H}v(h') \Q(h'|h)),\\
    c(h)+\lambda \sum_{h'\in S_H}v(h') \H(h'|h)
    \end{pmatrix},\\
    &~~~~V(h,K+1,m)= c(h)+\lambda \sum_{h'\in S_H}v(h') \H(h'|h),
\end{split}
\end{align}
where \begin{align*}
     v(h)=\sum_{k\in S_K} \left(\sum_{m\in S_M}\M(m)V(h,k,m)\right)  \K(k|h),~h\in S_H.
\end{align*}
Note that $V(H+1,k,m)=0,~\forall k\in S_K, ~m\in S_M$. Moreover, the sequence of functions $\{V_{n}\}_{n\geq 0}$ recursively defined by the value iteration procedure given by
\begin{align} \label{value-iteration-dis}
\begin{split}
& ~~~~V_{n+1}(h,k,m)\\&=\max\begin{pmatrix}
    (1-\F(h,k,m))r(h,k,m)+\F(h,k,m) (c(h)+\lambda \sum_{h'\in S_H}v_n(h') \Q(h'|h))\\
    c(h)+\lambda \sum_{h'\in S_H}v_n(h') \H(h'|h)
    \end{pmatrix},\\
&~~~~V_{n+1}(h,K+1,m)= c(h)+\lambda \sum_{h'\in S_H}v_n(h') \H(h'|h),
\end{split}
\end{align}
$\forall h\in S_H,~m \in S_M,~k<K+1$, converges pointwise to $V$, starting from any bounded function $V_0$, where 
\begin{align*}
     v_{n}(h)=\sum_{k\in S_K} \left(\sum_{m\in S_M}\M(m)V_n(h,k,m)\right)  \K(k|h),~h\in S_H.
\end{align*}
\end{defthm}

$v(h)$ can be interpreted as the expected total discounted reward when the patient state is $h$. Note that the optimal policy may not be unique, and denote $A^*(h,k,m)$ the set of optimal actions at state $(h,k,m)$.


\Cref{mono-hk-dis} provides sufficient conditions to guarantee that the value function $V(h,k,m)$ is nonincreasing in both $h$ and $k$. \Cref{mono-hk-dis} is intuitive: the patient overall benefit won't increase if the quality of the kidney offer or the patient state gets worse. First, we provide several intuitive assumptions and a preliminary result that are needed to establish \Cref{mono-hk-dis}.

\begin{defassump}\label{transplant-reward-dis}
$r(h,k,m)$ is nonincreasing in any component with the other two fixed.
\end{defassump}

Throughout the paper, we say that a multivariable function is monotone in some component if the function is monotone in that component with other components fixed. For example, $r(h,k,m)$ is nonincreasing in $h,k$, and $m$. \Cref{transplant-reward-dis} has an intuitive explanation that the reward for a successful transplantation does not increase if the patient state deteriorates and/or the kidney quality gets worse and/or the mismatch level increases.

\begin{defassump}\label{wait-reward-dis}
$c(h)$ is nonincreasing in $h$. 
\end{defassump}

\Cref{wait-reward-dis} is also intuitive: the intermediate reward for waiting does not increase if the patient state deteriorates.

\begin{defassump}\label{desenfail-dis}
$\F(h,k,m)$  is nondecreasing in any component with the other two fixed.
\end{defassump}

\Cref{desenfail-dis} can be interpreted as meaning that the probability of transplantation failure does not decrease if the patient state deteriorates and/or the kidney quality gets worse and/or the mismatch level increases. 



\begin{defdef}\label{ifr-dis}
For a time-homogeneous discrete-time Markov chain with state space $S=\{1,\cdots,n\}$, its transition probability function $P(\cdot|\cdot): S\times S\mapsto [0,1] $ is stochastically increasing \citep{smith2002structural} if the sequence of random variables with distribution functions $\{P(\cdot|k)\}_{k=1,\cdots,n}$ is in increasing stochastic order, which is defined as follows \citep{ross1996stochastic}: for random variables $X$ and $Y$, we say that $X \succeq_{st} Y$ if $\P(X>t) \geq \P(Y>t), \forall t$.
\end{defdef}

Specifically, $P(\cdot|\cdot)$ is stochastically increasing if $\sum_{j=k}^{n}P(j|i)$ is nondecreasing in $i$ for any $k=1,\cdots,n$, where $P(j|i)$ is the transition probability from state $i$ to $j$.

\begin{defassump}\label{markovifr-dis}
    Transition probability functions $\H$ and $\Q$ are stochastically increasing.
\end{defassump}

Some papers, e.g., \citet{alagoz2004optimal,alagoz2007determining,alagoz2007choosing}, refer to the stochastically increasing property as increasing failure rate (IFR). The stochastically increasing property has an intuitive explanation in the context of disease progression: the worse the patient state, the more likely the patient state is to become even worse.



\begin{defdef}\label{markov-st}
For a time-homogeneous discrete-time Markov chain with state space $S=\{1,\cdots,n\}$, we say that transition probability function $P(\cdot|\cdot): S\times S\mapsto [0,1]$ is stochastically greater than $Q(\cdot|\cdot): S\times S\mapsto [0,1]$, denoted by $P\succeq_{st} Q$, if
\begin{align*}
    \sum_{j=k}^n P(j|i) \geq \sum_{j=k}^n Q(j|i),~\forall i,k\in S.
\end{align*}
\end{defdef}

\begin{defassump}\label{diffifr-dis}
$\Q\succeq_{st} \H$. 
\end{defassump}

Note that larger patient state represents worse status. \Cref{diffifr-dis} implies that the patient state is less likely to become worse under transition function $\H$, compared with transition function $\Q$. \Cref{diffifr-dis} captures the negative impact of the transplantation failure: the patient state is more likely to become worse if they experience a failed transplantation.

\begin{defassump}\label{mono-k-assump}
For $k=1,\cdots,K,~h=1,\cdots,H$,
\begin{align}\label{rate-condition-dis}
      \K(k|h+1)\leq \K(k|h).
\end{align}
\end{defassump}

\Cref{mono-k-assump} posits that patients in better states are more likely to receive an offer, aligned with the current kidney allocation rule. For instance, patients with top $20\%$ EPTS scores are favored in the allocation of high-quality kidneys (i.e., those with KDPI less than 35.), and prioritization is given to patients who have longer waiting time. For more details, refer to Table 8-7 and Sections 8.2–8.4 in \citet{optnpolicy2023}. This differs from liver allocation, where patients in critical medical condition (i.e., those having high model for end-stage liver disease (MELD) scores) are prioritized. 

\begin{defthm}\label{mono-hk-dis}
Under \Cref{transplant-reward-dis,wait-reward-dis,desenfail-dis,markovifr-dis,diffifr-dis,mono-k-assump}, the following hold:
\begin{enumerate}
    \item $v(h)$ is nonincreasing in $h$;
    \item $V(h,k,m)$ is nonincreasing in $k$;
    \item $V(h,k,m)$ is nonincreasing in $h$.
\end{enumerate}
\end{defthm}

\begin{defrmk}
    \citet{smith2002structural} proves that the value function is decreasing when the one-stage reward function is decreasing for each action and the overall transition probability of the MDP is stochastically increasing (suitably defined for the partially ordered state space), for which \Cref{desenfail-dis,markovifr-dis,diffifr-dis,mono-k-assump} are neither necessary nor sufficient.
\end{defrmk}


Next, We will establish sufficient conditions to guarantee the existence of control limit-type optimal policies. First, we formally define a \textit{control limit policy}.

\begin{figure}[tb]
    \centering
\begin{tikzpicture}
\draw[->] (-1,0) -- (4.5,0) node [anchor=west] {$k$}; 
\draw[->] (0,-1) -- (0,4.5) node [anchor=south] {$h$};
\draw (0,0) .. controls (1,0) and (0,2) .. (4.5,4.5);
\node (X) at (1,3)  {$T$};
\node (Y) at (3,1)  {$W$};
\node (O) at (-0.5,-0.5)  {$O$};
\node (H) at (0,5.2)  {};
\end{tikzpicture}
\caption{\baselineskip8pt For an MDP with state space $S_H\times S_K$, an optimal policy such that states for which optimal actions are $W$ and $T$ are contained in two disjoint connected subsets.}\label{2acl}
\end{figure}

\begin{figure}[tb]
\centering
\begin{tikzpicture}[x  = {(-0.5cm,-0.5cm)},
                    y  = {(0.9659cm,-0.25882cm)},
                    z  = {(0cm,1cm)},
                    scale = 1.2,
                    color = {lightgray}]
\tikzset{facestyle/.style={fill=lightgray,draw=red,very thin,line join=round}}
\begin{scope}[canvas is xy plane at z=1]
  \path[facestyle] (0,0) rectangle (3,4);
\end{scope}
\begin{scope}[canvas is zx plane at y=2]
  \path[facestyle] (0,0) rectangle (2,3);
\end{scope}
\draw[very thin,red] (3,0,1) -- (3,3,1) ;
\draw[very thin,red,dashed] (0,2,1) -- (3,2,1) ;
\draw[very thin,red] (0,0,1) -- (0,3,1) ;
\node[black] (x) at (4,0,0)  {$h$};
\node[black] (z) at (0,0,3)  {$k$};
\node[black] (y) at (0,5,0)  {$m$};

\node[red] (t1) at (1.5,1.2,2.5)  {$T$};
\node[red] (t2) at (1.5,2.8,-0.4)  {$T$};
\node[green] (w1) at (1.5,0.6,-0.4)  {$W$};
\node[green] (w2) at (1.5,3.8,2.5)  {$W$};
     
\draw[very thin,black,->] (0,0,0) -- (3.5,0,0) ;
\draw[very thin,black,->] (0,0,0) -- (0,4.5,0) ;
\draw[very thin,black,->] (0,0,0) -- (0,0,2.5) ;
\end{tikzpicture}
\caption{\baselineskip8pt For an MDP with state space $S_H\times S_K\times S_M$, suppose that its unique optimal policy partitions $\R^3$ into four disjoint decision regions by a vertical plane and a horizontal plane. Both actions $W$ and $T$ are optimal over two disconnected regions. However, all three types of control limit optimal policies exist.}\label{3acl}
\end{figure}

\begin{defdef}\label{clpolicy}
Consider an MDP model with one-dimensional state space $S\subset \R$ and an action space $\A$. A policy $\pi:S\mapsto \A$ is called a \textit{control limit policy} if there exists a finite collection of intervals $\{I_i\}_{i=1}^n$ partitioning $\R$ and satisfying
\begin{enumerate}
    \item For any $a\in\A$, there exists at most one interval $I_i$ satisfying $\pi(s)=a,~\forall s\in I_i \bigcap S$;
    \item For any interval $I_i$, there exists $a\in \A$ such that $\pi(s)=a,~\forall s\in I_i\bigcap S$.
\end{enumerate}
Endpoints of intervals $\{I_i\}_{i=1}^n$ are called control limits.
\end{defdef}

The simplest form of control limit policy is the following, which partitions the state space into two regions:
\begin{align}
\begin{split}\label{contrlimit}
    \pi(s)=\begin{cases}
    a_1 &\text{if } s<s^*,\\
    a_2 &\text{if } s\geq s^*.
    \end{cases}   
\end{split}
\end{align}
The action to take depends only on whether the state $s$ is greater than or less than the control limit $s^*$, and solving the MDP problem boils down to finding the optimal threshold. Our setting necessitates a three-dimensional state vector, which requires an adjustment to \Cref{clpolicy}. Under suitable conditions, by fixing values of two state variables and projecting the state space onto the other dimension, we can establish optimal policies that take the form of a control limit policy (in one-dimension).

\begin{defdef}
A policy $\pi: S_H\times S_K\times S_M\mapsto \A$ is called a patient-based control limit policy if there exists a control limit function $H(k,m)$ such that for each $k\in S_K,~m\in S_M$, $\pi(h,k,m)=T$ if and only if the patient state $h>H(k,m)$ (or $h<H(k,m)$).

\end{defdef}

\begin{defdef}
A policy $\pi: S_H\times S_K\times S_M\mapsto \A$ is called a kidney-based control limit policy if there exists a control limit function $K(h,m)$ such that for each $h\in S_H,~m\in S_M$, $\pi(h,k,m)=T$ if and only if the kidney state $k<K(h,m)$ (or $k>K(h,m)$).
\end{defdef}

\begin{defdef}
A policy $\pi: S_H\times S_K\times S_M\mapsto \A$ is called a match-based control limit policy if there exists a control limit function $M(h,k)$ such that for each $h\in S_H,~k\in S_K$, $\pi(h,k,m)=T$ if and only if the mismatch level $m<M(h,k)$ (or $m>M(h,k)$).
\end{defdef}


\begin{defrmk}
For an MDP model with a two-dimensional state space $S_H\times S_K$, if both patient-based and kidney-based control limit optimal policies exist, it is easy to show that there exists an optimal policy such that states for which optimal actions are $W$ and $T$, respectively, are contained in two disjoint connected subsets of $\R^2_+$, as illustrated in \Cref{2acl}. There is only one decision boundary to determine. However, our MDP model has a three-dimensional state space. Existence of all three types of control limit policies does not guarantee that there exists an optimal policy that allows $\R^3_+$ to be partitioned into two connected decision regions. There are additional decision boundaries to identify, making both the computation and implementation of the policy more complicated. A counterexample is provided in \Cref{3acl}. \citet{ren2022review} provide an example showing that the expansion of the action space has a similar effect. In general, the optimal policies may take a more complicated form if the size of either the state space or the action space or both increase.
\end{defrmk}

We shall see later that all three types of control limit optimal policies are equivalent if all of them exist, i.e., once we derive one control limit function, the other two can be obtained by taking the inverse of the derived one. 

The control limit-type policy is a specific case of a monotone policy. In their work, \citet{serfozo1976monotone} establishes sufficient conditions for the existence of a monotone optimal policy within the context of a discrete-time MDP, where the state space $S$ is assumed to be partially ordered, and the action space $A$ is assumed to be a compact subset of the real line. Define the Q-function
\begin{align}\label{Q-function}
\begin{split}
    \overline Q(h,k,m,a) =\begin{cases}
        (1-\F(h,k,m))r(h,k,m)\\+\F(h,k,m) (c(h)+\lambda \sum_{h'\in S_H}v(h') \Q(h'|h))
        &\text{if }a=T,\\
        c(h)+\lambda \sum_{h'\in S_H}v(h') \H(h'|h) &\text{if }a=W.
    \end{cases}    
\end{split}    
\end{align}
\citet{serfozo1976monotone} proves the existence of a monotone optimal policy when the Q-function is submodular on $S \times A$, where submodularity is defined as follows.
\begin{defdef}
    Let $X$ and $Y$ be partially ordered sets and $f(x,y)$ be a real-valued function on $X\times Y$. We say that $f$ is submodular if for $x_1\geq x_2$ in $X$ and $y_1\geq y_2$ in $Y$,
    \begin{align*}
        f(x_1,y_1) + f(x_2,y_2) \leq f(x_1,y_2) +f (x_2,y_1).
    \end{align*}
\end{defdef}
Applying the result of \citet{serfozo1976monotone}, it is evident that by selecting any state variable and fixing the other two, if the Q-function is submodular as a function of that variable and the action, a control limit-type optimal policy for that variable can be established (although the action space is unordered, we could artificially assign an order to it, such as $T>W$ or $T<W$). Specifically,
\begin{itemize}
    \item The patient-based control limit optimal policy exists if $\overline Q(h+1,k,m,T)-\overline Q(h,k,m,T)>\overline Q(h+1,k,m,W)-\overline Q(h,k,m,W),~\forall h$.
    \item The kidney-based control limit optimal policy exists if $\overline Q(h,k+1,m,T)-\overline Q(h,k,m,T)<\overline Q(h,k+1,m,W)-\overline Q(h,k,m,W),~\forall k$.
    \item The match-based control limit optimal policy exists if $\overline Q(h,k,m+1,T)-\overline Q(h,k,m,T)<\overline Q(h,k,m+1,W)-\overline Q(h,k,m,W),~\forall m$.
\end{itemize}
The first condition implies that for a patient-based control limit policy to be optimal, as a patient’s health deteriorates  (i.e., as $h$ increases), the reduction in benefit from waiting must exceed the reduction in benefit from immediate transplantation. In practice, this situation may arise when a patient is in poor health, where the decline in quality of life on dialysis or the increased risk of death during waiting makes immediate transplantation more preferable as the patient state worsens. The other two conditions can be explained in a similar manner. \Cref{match-based-dis,kidney-based-dis,health-based-dis} will provide sufficient conditions to establish the aforementioned conditions, respectively.

With \Cref{transplant-reward-dis,wait-reward-dis,desenfail-dis,markovifr-dis,diffifr-dis,mono-k-assump}, we are able to prove \Cref{match-based-dis,kidney-based-dis}, which establish match-based and kidney-based control limit optimal policies, respectively.

\begin{defthm} \label{match-based-dis}
Under \Cref{transplant-reward-dis,wait-reward-dis,desenfail-dis,markovifr-dis,diffifr-dis,mono-k-assump}, there exists a \textit{match-based} control limit optimal policy, i.e., there exists a control limit function $M^*(h,k)$, such that for any fixed $h$ and $k$, it is optimal to accept the kidney offer if and only if the mismatch level $m<M^*(h,k)$
\end{defthm}

\begin{defthm} \label{kidney-based-dis}
Under \Cref{transplant-reward-dis,wait-reward-dis,desenfail-dis,markovifr-dis,diffifr-dis,mono-k-assump}, there exists a \textit{kidney-based} control limit optimal policy, i.e., there exists a control limit function $K^*(h,m)$, such that for any fixed $h$ and $m$, it is optimal to accept the kidney offer if and only if the kidney state $k<K^*(h,m)$.
\end{defthm}

\begin{defrmk}
Although the optimal policy may not be unique, ``if and only if" in the theorem guarantees that there exists a unique match-based (and kidney-based) control limit optimal policy. 
\end{defrmk}

Using \Cref{match-based-dis}, it can be easily shown that $V(h,k,m)$ is nonincreasing in $m$, i.e., the value function doesn't increase if the mismatch level increases.

\begin{defcoro}\label{mono-m-dis}
Under \Cref{transplant-reward-dis,wait-reward-dis,desenfail-dis,markovifr-dis,diffifr-dis,mono-k-assump}, $V(h,k,m)$ is nonincreasing in $m$.
\end{defcoro}

Both \Cref{match-based-dis,kidney-based-dis} are intuitive: the decision maker should accept kidney offers that are of sufficiently good quality and/or low mismatch level. Once we establish both control limit optimal policies, we can easily show the (partial) monotonicity, as well as invertibility of control limit functions.

\begin{defcoro}\label{mono-limit-1-dis}
Under \Cref{transplant-reward-dis,wait-reward-dis,desenfail-dis,markovifr-dis,diffifr-dis,mono-k-assump}, the following hold:
\begin{enumerate}
    \item $K^*(h,m)$ is nonincreasing in $m$.
    \item $M^*(h,k)$ is nonincreasing in $k$.
    \item Define $K^{-M}(h,k):=\min \{m\in S_M \ | \ k\geq K^*(h,m)\}$. Then, $M^*(h,k)=K^{-M}(h,k)$.
    \item Define $M^{-K}(h,m):=\min\{ k\in S_K \ | \ m\geq M^*(h,k) \}$. Then, $K^*(h,m)=M^{-K}(h,m)$.
\end{enumerate}
\end{defcoro}

To show \Cref{health-based-dis}, the existence of a patient-based control limit optimal policy, we need several additional assumptions.

\begin{defassump}\label{h-based-con1}
For $h=1,\cdots,H \text{ and } h_0=h+1,\cdots,H$,
    \begin{align}
        \sum_{h'=h_0}^H \H(h'|h) \leq \sum_{h'=h_0}^H \H(h'|h+1).
    \end{align}
\end{defassump}

The interpretation of \Cref{h-based-con1} is similar to the stochastically increasing property, but \Cref{h-based-con1} is neither a sufficient nor a necessary condition for the stochastically increasing property of $\H$.

\begin{defassump}\label{h-based-con2}
 For $h=1,\cdots,H-1$, $k\leq K,~ m\in S_M$
    \begin{align}
    \begin{split}
        &~~~~\frac{\E g((h,k,m),T)-\E g((h+1,k,m),T)}{\E g((h+1,k,m),T)}\\ &\leq (1-\F(h,k,m)) \lambda(\H(H+1|h+1)-\H(H+1|h)).
    \end{split}
    \end{align}
\end{defassump}

\Cref{h-based-con2} has an intuitive explanation that, as the patient state becomes worse, the increment of the probability of death during waiting is greater than the \textit{marginal} reduction in the expectation of the one-step reward for choosing $T$. Conditions similar to \Cref{h-based-con2} have been verified using real data in liver transplant studies \citep{alagoz2004optimal,alagoz2007determining}.

\begin{defassump}\label{h-based-con3}
    $\Q-\H$ is stochastically decreasing, i.e., for $h_0=1,\cdots,H+1$ and $h=1,\cdots,H$,
    \begin{align}
         \sum_{h'=h_0}^{H+1} \Q(h'|h+1)-\H(h'|h+1) \leq \sum_{h'=h_0}^{H+1}\Q(h'|h)-\H(h'|h).
    \end{align}
\end{defassump}


\Cref{h-based-con3} states that as patient state $h$ increases, the ``distance'' between distributions $\H(\cdot|h)$ and $\Q(\cdot|h)$ decreases. In particular, \Cref{h-based-con3} implies that transplantation failure has a larger impact on a healthier patient, e.g., an unsuccessful transplant will result in a substantial increase in EPTS score for a low-EPTS patient, as shown in \Cref{EPTS-evolve,EPTS-fail}.


\begin{defthm}\label{health-based-dis}
Under \Cref{transplant-reward-dis,wait-reward-dis,desenfail-dis,markovifr-dis,diffifr-dis,mono-k-assump,h-based-con1,h-based-con2,h-based-con3}, there exists a \textit{patient-based} control limit optimal policy, i.e., there exists an optimal control limit function $H^*(k,m)$, such that for any fixed $k$ and $m$, it is optimal to accept the kidney offer if and only if the patient state $h>H^*(k,m)$.
\end{defthm}

\Cref{health-based-dis} has an intuitive explanation: the decision maker should accept a kidney offer if the patient state is worse than some threshold. With \Cref{health-based-dis}, we are able to derive more monotonicity and invertibility results of control limit functions similar to \Cref{mono-limit-1-dis}.

\begin{defrmk}
    Under appropriate conditions, the existence of a monotone optimal policy has been established for various MDP types, including regular MDPs \citep{serfozo1976monotone,puterman2014markov,flores2007monotonicity}, risk-sensitive MDPs \citep{avila1998controlled}, and partially observed MDPs \citep{lovejoy1987some,miehling2020monotonicity}. These conditions typically involve assumptions about the stochastic monotonicity of the overall transition probability and the submodularity of the one-period reward function. In contrast, \Cref{kidney-based-dis,match-based-dis,health-based-dis} require only stochastic monotonicity in each of the individual dimensions of the state transition, aligning with the metatheorems in \citet{oh2016characterizing}, which provide a framework for establishing threshold policies in optimal stopping problems with partially monotone state transitions.
\end{defrmk}


\begin{defcoro}\label{mono-limit-2-dis}
Under \Cref{transplant-reward-dis,wait-reward-dis,desenfail-dis,markovifr-dis,diffifr-dis,mono-k-assump,h-based-con1,h-based-con2,h-based-con3}, the following hold:
\begin{enumerate}
    \item $K^*(h,m)$ is nondecreasing in $h$.
    \item $M^*(h,k)$ is nondecreasing in $h$.
    \item $H^*(k,m)$ is nondecreasing in $k$ and $m$.
    \item Define $K^{-H}(k,m):=\max \{h\in S_H\ | \ k\geq K^*(h,m)\}$. Then, $H^*(k,m)=K^{-H}(k,m)$.
    \item Define $M^{-H}(k,m):=\max\{ h\in S_H \ | \ m\geq M^*(h,k) \}$. Then, $H^*(k,m)=M^{-H}(k,m)$.
    \item Define $H^{-M}(h,k):=\min\{ m\in S_M \ | \ h\leq H^*(k,m) \}$. Then, $M^*(h,k)=H^{-M}(h,k)$.
    \item Define $H^{-K}(h,m):=\min\{ k\in S_K \ | \ h\leq H^*(k,m) \}$. Then, $K^*(h,m)=H^{-K}(h,m)$.
\end{enumerate}
\end{defcoro}

The proof of \Cref{mono-limit-2-dis} is omitted, as it is straightforward and exactly the same as that of \Cref{mono-limit-1-dis}. \Cref{mono-limit-2-dis} together with \Cref{mono-limit-1-dis} shows that the three types of control limit optimal policies are equivalent to each other: if all three types of control limit optimal policies exist, once we obtain any one of them, we can easily obtain the other two by invertibility.

\Cref{diff-avai} considers two patients with identical state transition probability functions but different probabilities for kidney offers. For example, compared with insensitive patients, it is much harder for highly sensitized patients to find a compatible donor. If patient 1 has a higher chance to receive a kidney offer than patient 2, then the value function of patient 1 dominates the value function of patient 2. Recall that a larger value of $k$ indicates worse quality and $k=K + 1$ means no kidney offer is available.

\begin{defthm}\label{diff-avai}
Let $\Pi_1$ and $\Pi_2$ be two MDPs, where the distributions of the kidney state are $\K_1$ and $\K_2$, respectively. Suppose that $\K_2 \succeq_{st} \K_1$. Let $V^1$ and $V^2$ be the value functions of $\Pi_1$ and $\Pi_2$, respectively. If $\Pi_1$ and $\Pi_2$ have the same reward functions $c$ and $r$, probability function of transplantation failure $\F$, pmf of mismatch level $\M$, and patient state transition probability functions $\H$ and $\Q$, 
then $ V^1(h,k,m)\geq V^2(h,k,m)$ for all $h\in S_H,k\in S_K,m\in S_M$.
\end{defthm}

\Cref{diff-trans} provides a similar result, which considers two patients with identical probabilities for kidney offers but different patient state transition probability functions. If the health of patient 1 is less likely to become worse, then the value function of patient 1 dominates the value function of patient 2. Recall that a larger value of $h$ implies worse patient state and $h=H+1$ represents death.

\begin{defthm}\label{diff-trans}
Let $\Pi_1$ and $\Pi_2$ be two MDPs with patient state transition probability functions $(\H_1,\Q_1)$ and $(\H_2,\Q_2)$, respectively. Suppose that $\H_2\succeq_{st} \H_1,~\Q_2\succeq_{st} \Q_1$. Let $V^1$ and $V^2$ be value functions of $\Pi_1$ and $\Pi_2$, respectively. If $\Pi_1$ and $\Pi_2$ have the same reward functions $c$ and $r$, probability function of transplantation failure $\F$, pmf of mismatch level $\M$, and distribution of kidney state $\K$, 
then $V^1(h,k,m)\geq V^2(h,k,m)$ for all $h\in S_H,k\in S_K,m\in S_M$.
\end{defthm}

%

\section{Numerical Experiments}\label{sec4}
In this section, we use numerical experiments to demonstrate the importance of modeling both kidney quality and compatibility and the impact of allowing the desensitization therapy option. We set parameters based on a recent OPTN data report \citep{unos2022}. Next, we compute the optimal policy derived in \Cref{sec3} and show how its behavior and performance vary under different kidney quality and mismatch levels. Then, we compare the optimal policy with another policy that doesn't include compatibility as a state variable or retransplantation. In particular, \textit{the optimal policy performs much better at low mismatch levels}. Moreover, we provide an example where \Cref{h-based-con2}, part of the sufficient condition for \Cref{health-based-dis} to hold, is violated and the patient-based control limit optimal policy doesn't exist. We observe that retransplantation is rare under these parameter settings due to the low likelihood of receiving a kidney offer and transplantation failure. To evaluate the effects of compatibility and retransplantation separately, we increase the chances of receiving a kidney offer and transplantation failure, and conduct an additional experiment where the mismatch level is hidden, but retransplantation is included.

We consider a $70-$year old patient who starts dialysis at the beginning of the decision process and doesn't have diabetes or a prior organ transplant. We begin by describing how state variables, state space and reward functions are defined. The \textit{decision period} is six months. The \textit{patient state} $h$, taking values in $S_H=\{1,\cdots,17\}$, is defined based on the EPTS score \citep{unosEPTS,bae2019cansurvival}, one of the most commonly used measures in evaluating the patient expected post-transplantation survival time. The patient state transition law is straightforward, because the EPTS score is a function of age once the patient's diabetes state, time to start dialysis, and number of prior organ transplantations are specified. We only consider deceased kidney donors and use the KDPI score to represent the \textit{kidney state} $k$. Both EPTS and KDPI scores are used in the current kidney allocation system. Since a recent OPTN data report \citep{unos2022} categorizes kidneys into four groups based on their KDPI ranges, we define four types of kidneys, i.e., $S_k=\{1,\cdots,5\}$ where $k=5$ represents that the kidney offer is not available. We use the degree of HLA mismatch to represent the \textit{mismatch level} $m$. The number of HLA antigen mismatches of a donor-recipient pair ranges from $0$ to $6$, so we set $S_M=\{1,\cdots,7\}$. We assume that kidney states $\{k_n\}_{n\geq 1}$ form an i.i.d. sequence of random variables, independent of both patient state and mismatch level. If the decision maker chooses to wait, the patient accrues an intermediate reward for being alive for half a year (the decision period), i.e., $c(h)=0.5,~\forall h$. If the patient undergoes a successful transplantation, they receive terminal reward $r(h,k,m)$ equal to the expected post-transplant lifetime. 

We consider two experiments. The first setting models many of the parameter values for a typical 70-year old patient with ESKD. The second setting is similar, with the exception that $\H(H+1|h)$, the probability of death in state $h$, is perturbed such that \Cref{h-based-con2} is no longer satisfied, so that \Cref{health-based-dis} no longer applies. Value iteration is used to solved both MDPs. 


We now describe some of the model parameters, with full details of the remaining parameter settings provided in \Cref{appendb}. As mentioned earlier, the patient state $h$ is defined based on the EPTS score, a deterministic function of age (or equivalently, the decision period) in this case. If the patient is in state $h,~h<H$ and chooses $W$, they either die or transition to state $h+1$ at the next epoch. We further assume that $\H(H+1|h)$ is an increasing affine function of $h$. Consequently, for $h< H$,
\begin{align}\label{patienthealth}
\begin{split}
    &\H(h'|h) =
    \begin{cases}
    a + b (h-1) & \text{if } h'=H+1,\\
    1-(a + b (h-1)) & \text{if } h'=h+1,\\
    0 &\text{otherwise},
    \end{cases}\\
    &\H(h'|H) =
    \begin{cases}
    a + b (H-1) & \text{if } h'=H+1,\\
    1-(a + b (H-1)) & \text{if } h'=H,\\
    0 &\text{otherwise},
    \end{cases} 
\end{split}
\end{align}
where $a=0.01$ in both experiments, $b=0.007$ in \hyperlink{exper1}{Experiment 1} and $0.006$ in \hyperlink{exper2}{Experiment 2}. The definition of the transition function $\Q$ is similar. Other parameters are the same in both experiments. We set the discount factor $\lambda = 0.99$. The following parameters are assigned according to \citet{unos2022}. The pmf of the kidney state $k$, from $k=1$ to $5$, is $(0.0491,0.0323,0.1206,0.0347,0.7653)$. The pmf of the mismatch level $m$, from $m=1$ to $7$, is $(0.0492,0.0104,0.0192,0.1437,0.2806,0.3254,0.1414)$. Because patients of age $70$ and older have EPTS scores greater than $20$ and \citet{unos2022} doesn't distinguish patients with EPTS scores greater than $20$, we assume in this section that \textit{the probability of a transplantation failure} depends only on $k$ and $m$, not on $h$, and denote it by $\F(k,m)$. The values of $\F(k,m)$ are summarized in \Cref{succ-trans-1}.

\begin{table}[t]
\tbl{The probability of a transplantation failure $\F(k,m)$.}
{
\setlength\extrarowheight{1pt}
\begin{tabular}{|c|c|c|}
\hline
$\F(k,m)$ & $m=1$ & $m>1$ \\ \hline
$k=1$       & 0.017            & 0.041          \\ \hline
$k=2$      & 0.037            & 0.061          \\ \hline
$k=3$      & 0.047            & 0.071          \\ \hline
$k=4$     & 0.073           & 0.095         \\ \hline
\end{tabular}
}
\label{succ-trans-1}
\end{table}

The values of the post-transplant reward $r(h,k,m)$ are calculated according to \citet{bae2019can} and provided in \Cref{ptreward} for $m=1$ and $7$, with the rest provided in \Cref{appendb}.

\begingroup
\renewcommand*{\arraystretch}{0.5}
\begin{align}\label{ptreward}
\begin{split}
        r(h,k,1)=\begin{bmatrix}
    12    & 11  & 10 & 8.5  \\
11  & 11  & 9.5 & 8.2  \\
9.9    & 9.7  & 9  & 7.9  \\
9.6 & 9.4 & 8.8 & 7.7     \\
9.3  & 9.1 & 8.5 & 7.6   \\
8.9 & 8.8  & 8.3  & 7.4  \\
8.7  & 8.5  & 8 & 7.2  \\
8.5  & 8.4  & 7.8 & 7.1  \\
8.3 & 8.1 & 7.7 & 6.9  \\
8.1 & 8  & 7.5 & 6.8  \\
8.1 & 8  & 7.5 & 6.8  \\
8 & 7.9  & 7.4  & 6.7   \\ 
7.8  & 7.7  & 7.3 & 6.6   \\
7.7  & 7.6  & 7.2 & 6.6   \\
7.7  & 7.6  & 7.1 & 6.5   \\
7.6  & 7.5  & 7.1  & 6.5
    \end{bmatrix},~
    r(h,k,7)=\begin{bmatrix}
6    & 5.9  & 5.8 & 5.5  \\
5.9  & 5.9  & 5.8 & 55  \\
5.8    & 5.8  & 5.7  & 5.4  \\ 
5.8 & 5.7 & 5.6 & 5.3     \\
5.7 & 5.7 & 5.6 & 5.3     \\ 
5.6  & 5.6 & 5.5 & 5.2   \\
5.6 & 5.6  & 5.4  & 5.1  \\
5.5  & 5.5  & 5.3 & 5.1  \\ 
5.5  & 5.4  & 5.3 & 5  \\ 
5.4 & 5.4 & 5.3 & 5  \\ 
5.4 & 5.4 & 5.3 & 5  \\
5.4 & 5.4  & 5.2 & 4.9  \\ 
5.4 & 5.3  & 5.2  & 4.9   \\
5.3  & 5.3  & 5.1 & 4.8   \\ 
5.3  & 5.2  & 5.1 & 4.8   \\
5.3  & 5.2  & 5.1 & 4.8 
    \end{bmatrix}.
\end{split}
\end{align}
\endgroup

\subsection*{\hypertarget{exper1}{Experiment 1}}
\begin{figure}[H]
    \centering
    \subfloat[Given $(h,k)$, the optimal action is to wait ($W$) if and only if the mismatch level $m$ is above the surface.]{%
    \resizebox*{6.9cm}{!}{\includegraphics{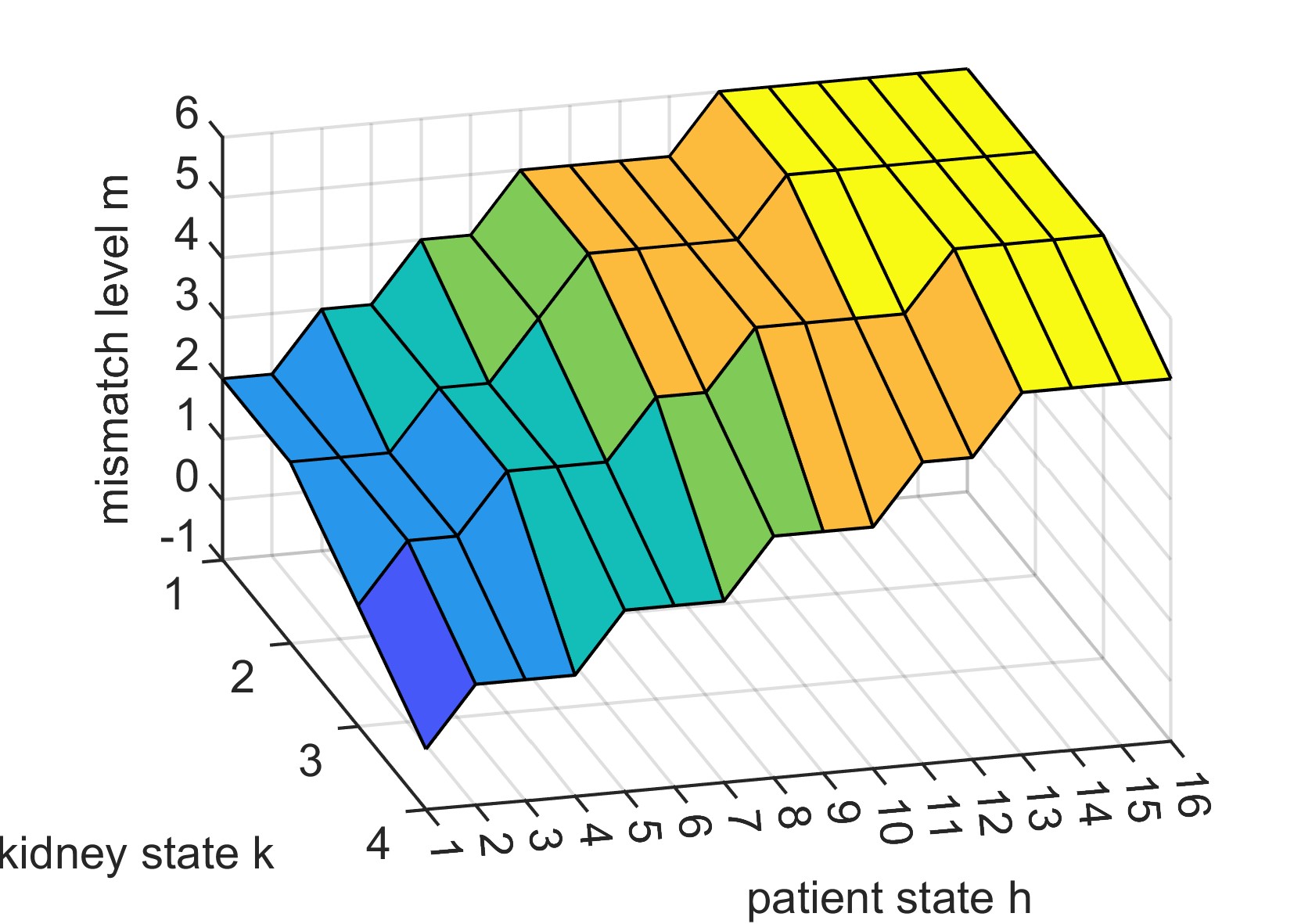}}}\hspace{5pt}
    \subfloat[Given patient state $h$, the optimal action is to wait ($W$) if and only if the kidney state $k$ is above the corresponding curve.]{%
    \resizebox*{5cm}{!}{\includegraphics{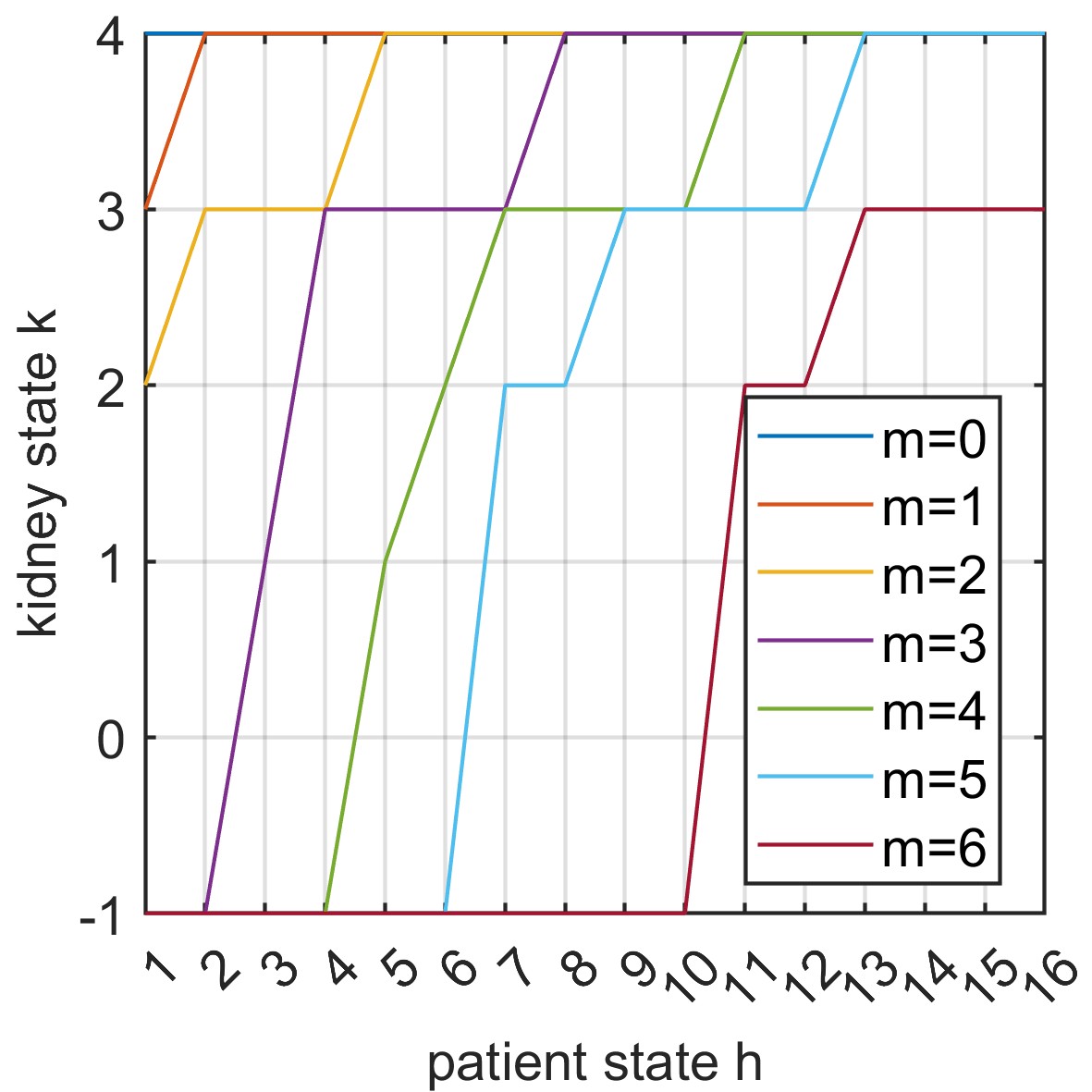}}}\hspace{5pt}
    \caption{The optimal policy $d_1$.} \label{3d-1-ex1}
\end{figure}

The optimal policy $d_1$ is given in \Cref{3d-1-ex1}. All three types of control limit optimal policies exist. For each mismatch level, we plot the projection of the optimal policy on the $h-k$ plane. 

\begin{figure}[h]
    \centering
    \subfloat[$m=1$.]{%
    \resizebox*{5cm}{!}{\includegraphics{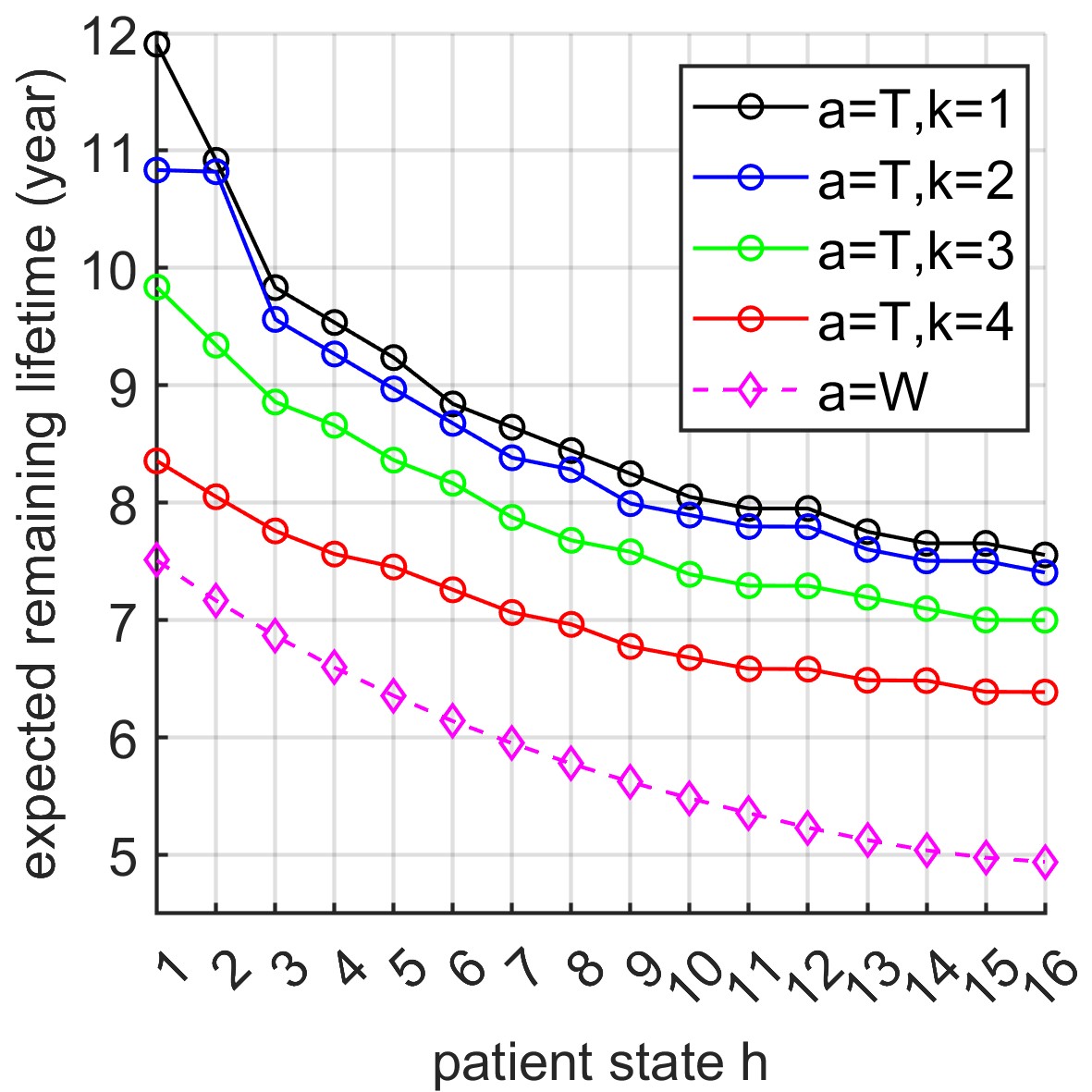}}}\hspace{5pt}
    \subfloat[$m=2$.]{%
    \resizebox*{5cm}{!}{\includegraphics{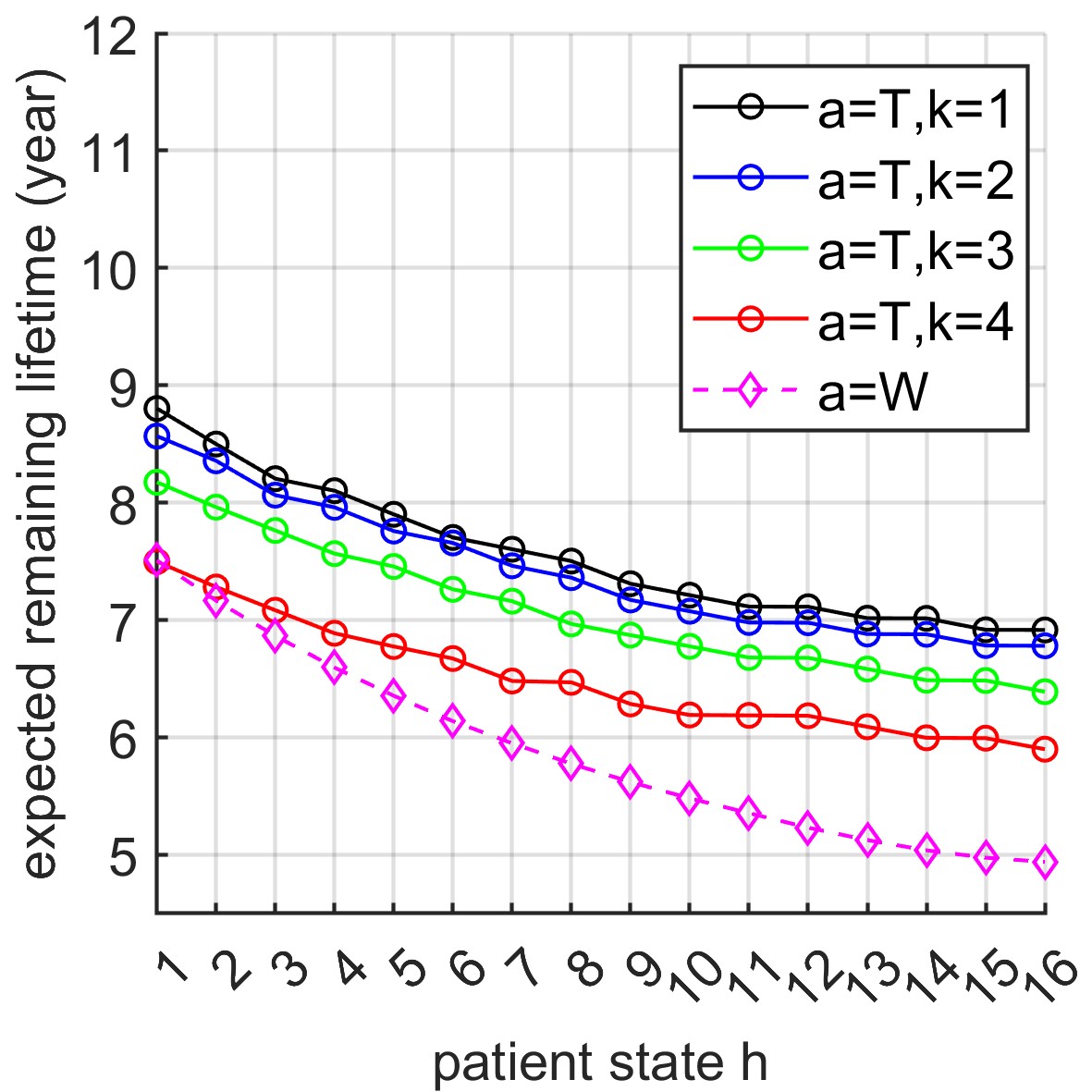}}}\hspace{5pt}
    \subfloat[$m=4$]{%
    \resizebox*{5cm}{!}{\includegraphics{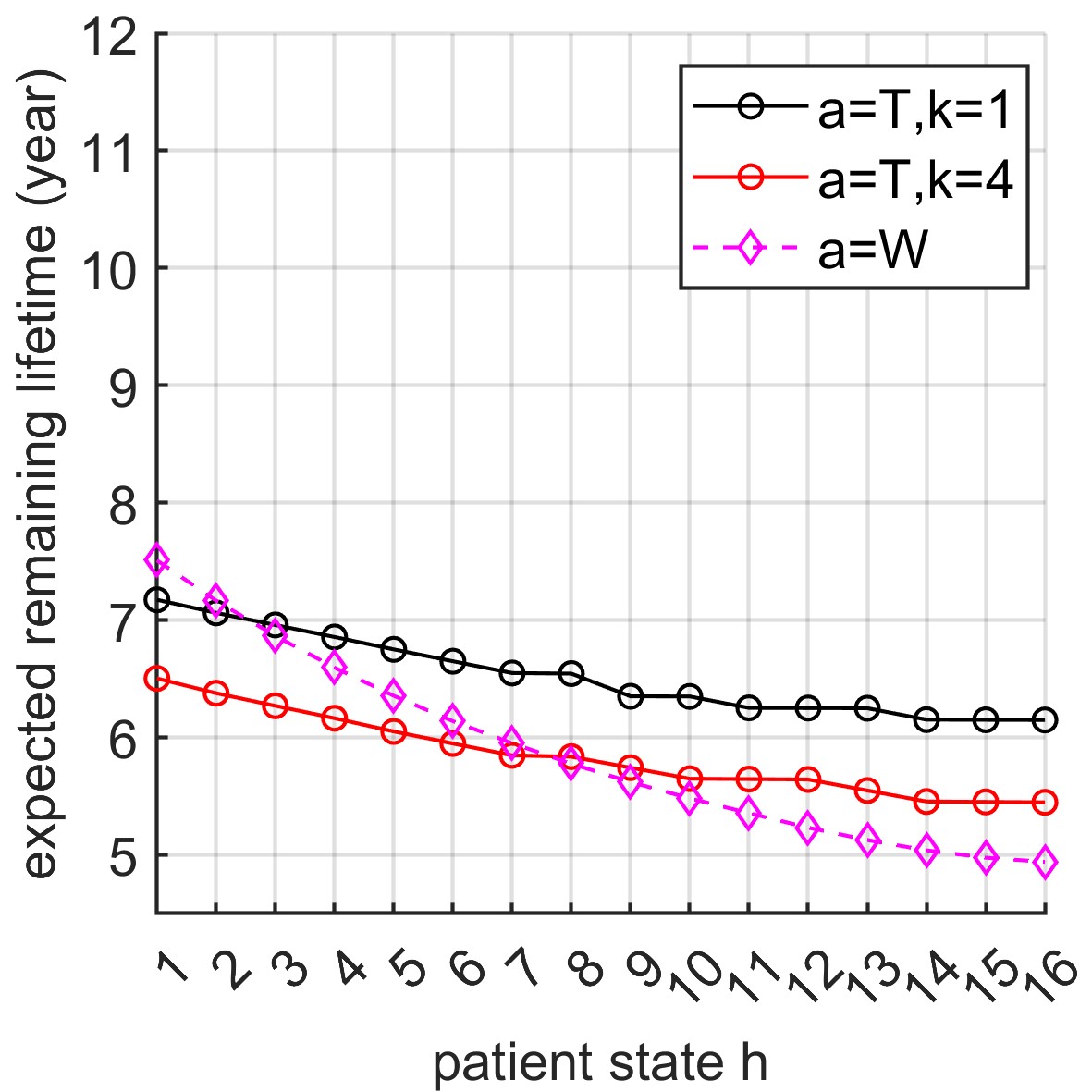}}}\hspace{5pt}
    \subfloat[$m=7$.]{%
    \resizebox*{5cm}{!}{\includegraphics{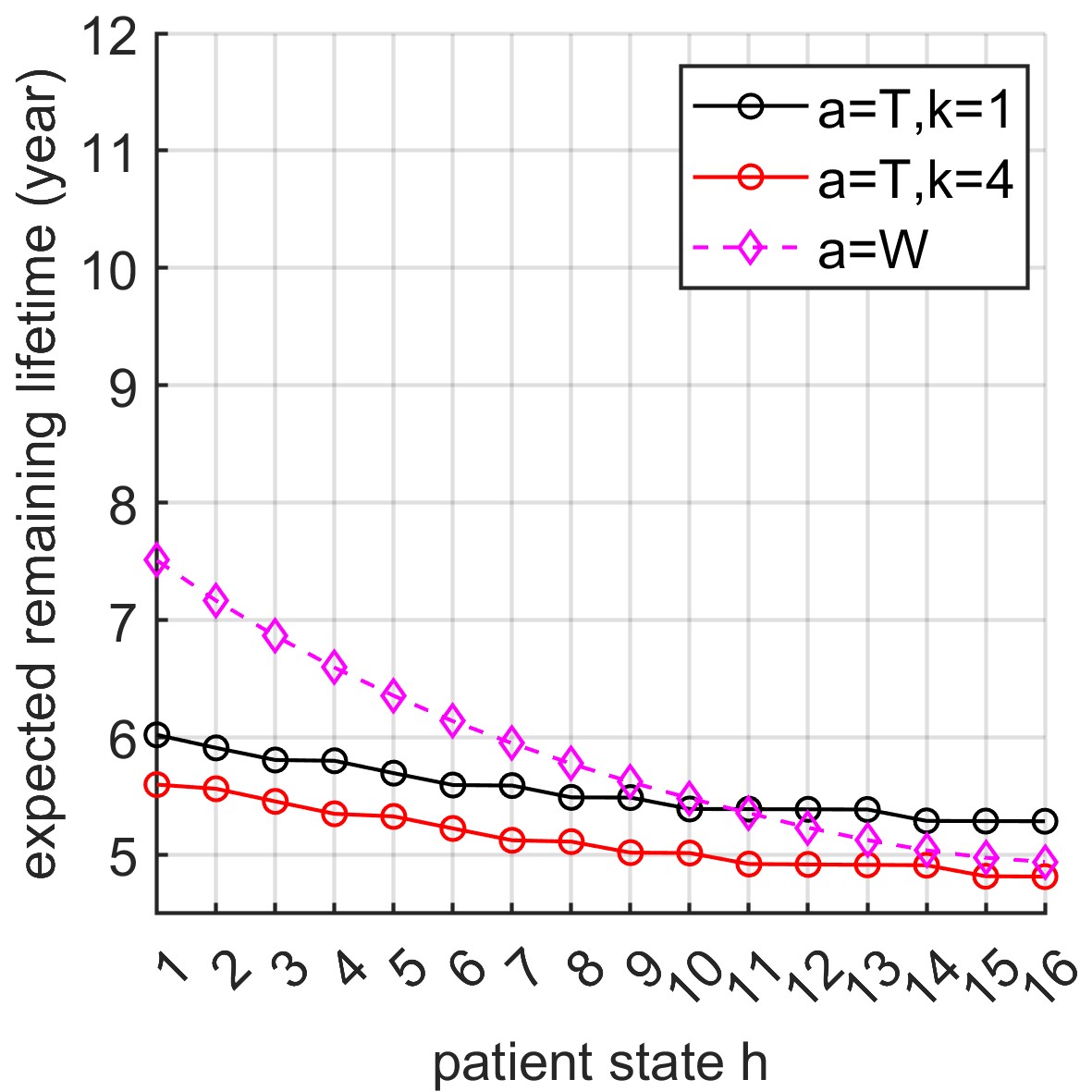}}}\hspace{5pt}
    \caption{Q-function for different mismatch levels.} \label{q-ex1}
\end{figure}

The Q-function is the expected total discounted reward starting from a given state, taking a given action, and following the optimal policy thereafter. We plot in \Cref{q-ex1} the Q-function, given by \Cref{Q-function}.

Note that the value of the Q-function depends only on the patient state $h$ if the action is to wait. When the mismatch level is low, it is more beneficial to accept an offer unless it is of very low quality. As the mismatch level increases, the optimal action is to wait and then switch to transplant if the patient state $h$ is below some control limit (i.e., the patient health is worse than some threshold). For fixed mismatch level, as the patient state $h$ worsens, the optimal policy tends to accept low-quality kidneys (i.e., those with high KDPI) that are rejected at lower $h$'s. At high mismatch levels, the decision maker should choose transplant only when the patient is in severe health status (i.e., $h$ is sufficiently large).

\clearpage
\begin{figure}[H]
    \centering
    \resizebox*{5cm}{!}{\includegraphics{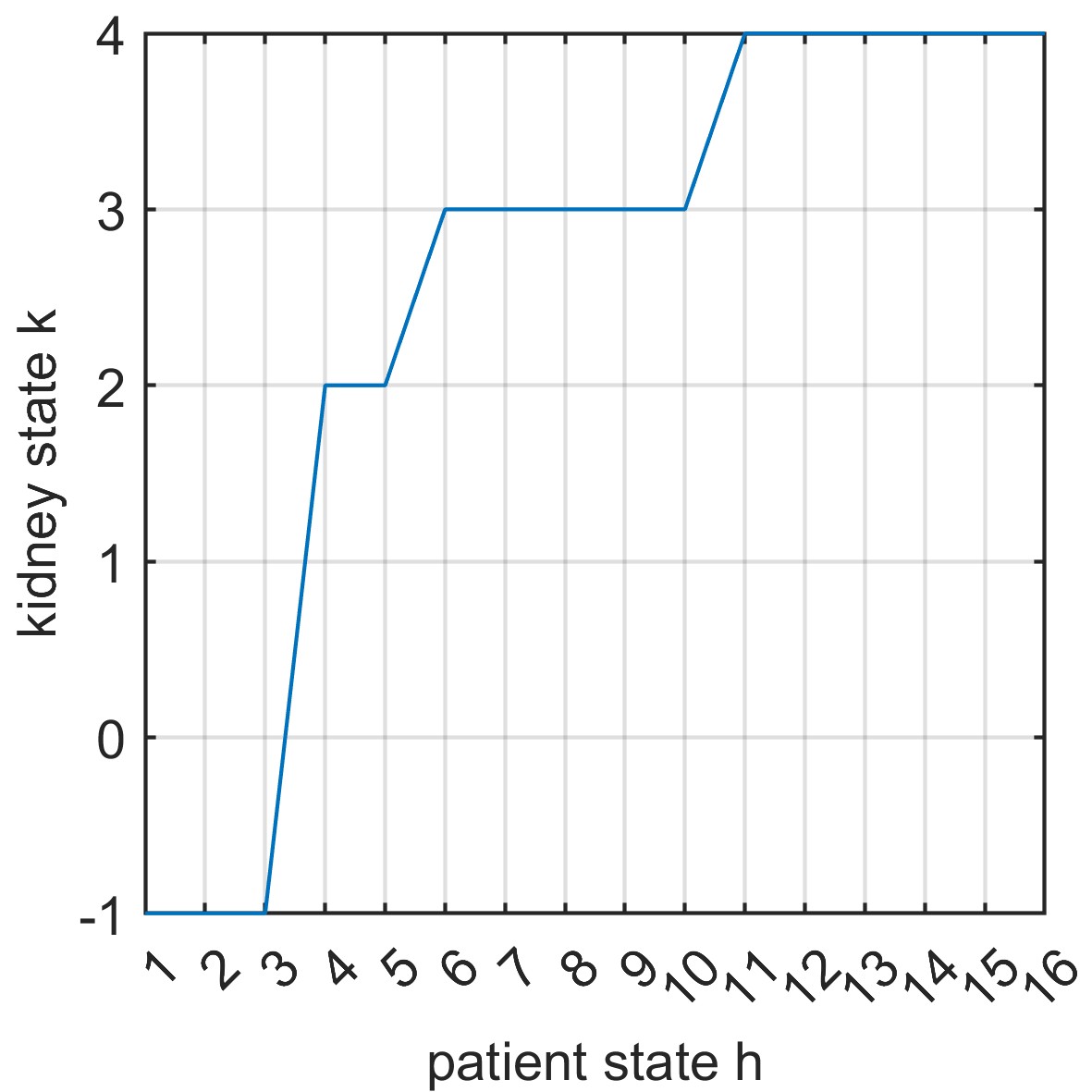}}
    \caption{Policy $q_1$: given patient state $h$, the optimal action is to wait ($W$) if and only if the kidney state $k$ is above the curve.}
    \label{2d-2-ex1}
    \vspace{-5mm}
\end{figure}

\begin{figure}[H]
    \centering
    \subfloat[$m=1$.]{%
    \resizebox*{5cm}{!}{\includegraphics{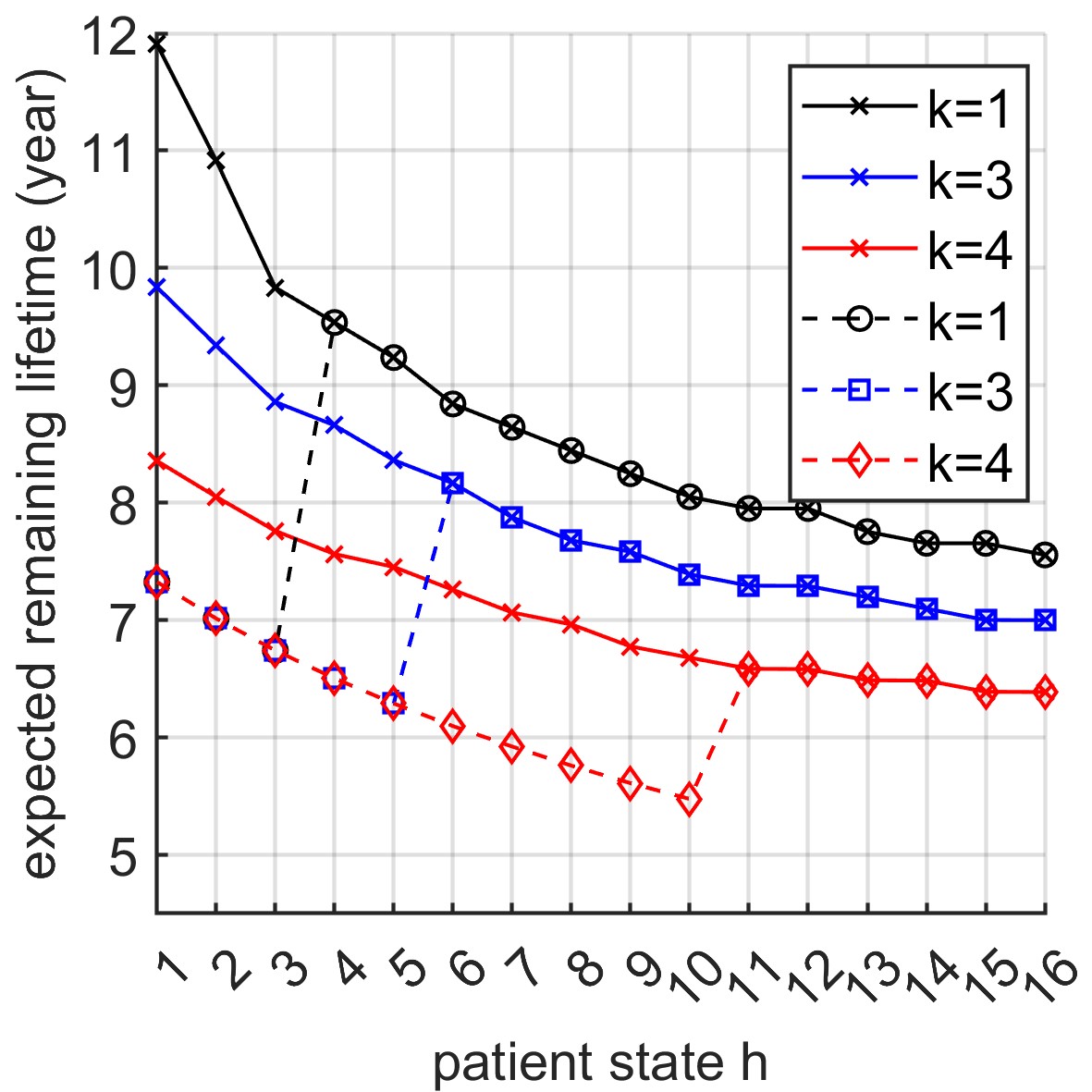}}}\hspace{5pt}
    \subfloat[$m=2$.]{%
    \resizebox*{5cm}{!}{\includegraphics{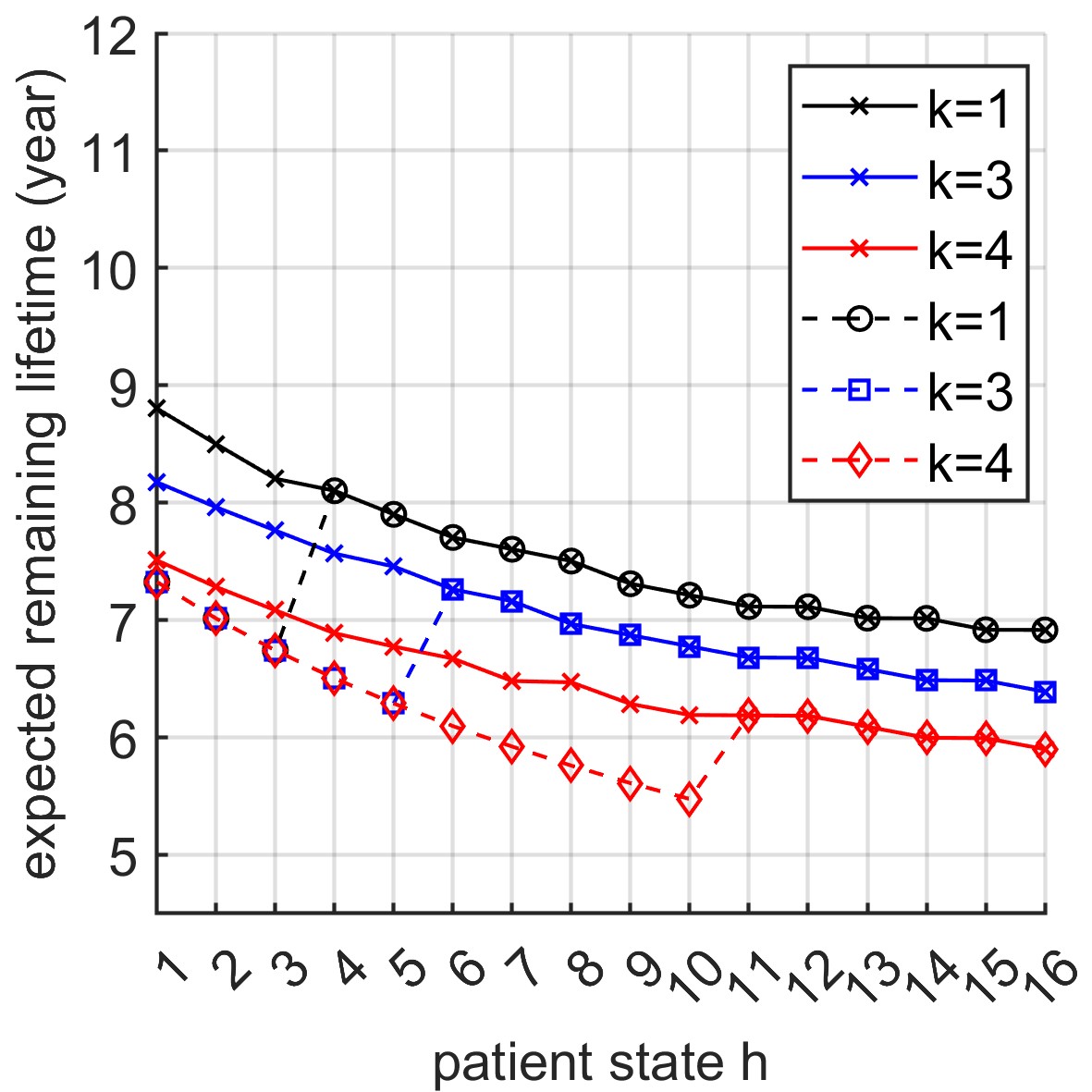}}}\hspace{5pt}
    \subfloat[$m=4$]{%
    \resizebox*{5cm}{!}{\includegraphics{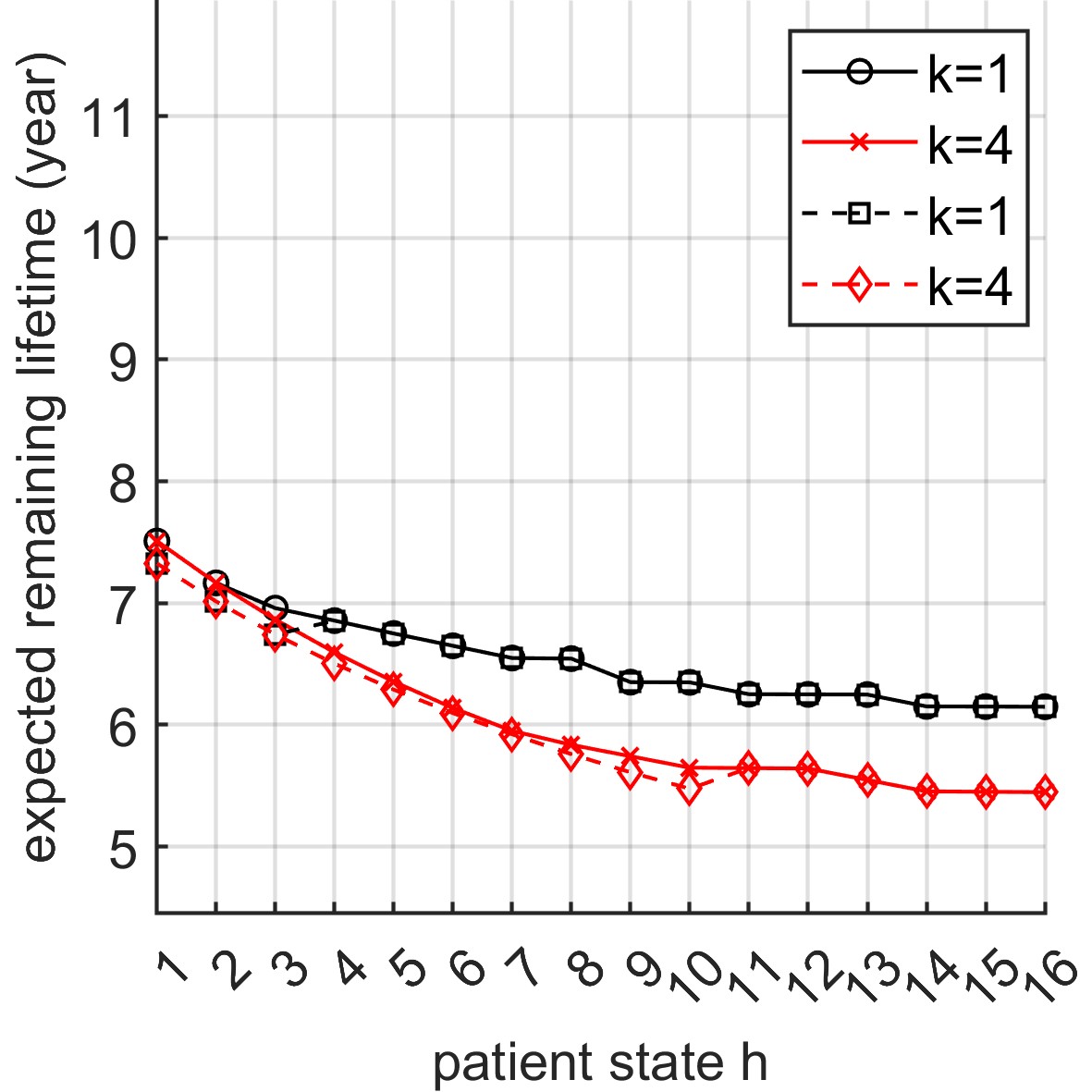}}}\hspace{5pt}
    \subfloat[$m=7$.]{%
    \resizebox*{5cm}{!}{\includegraphics{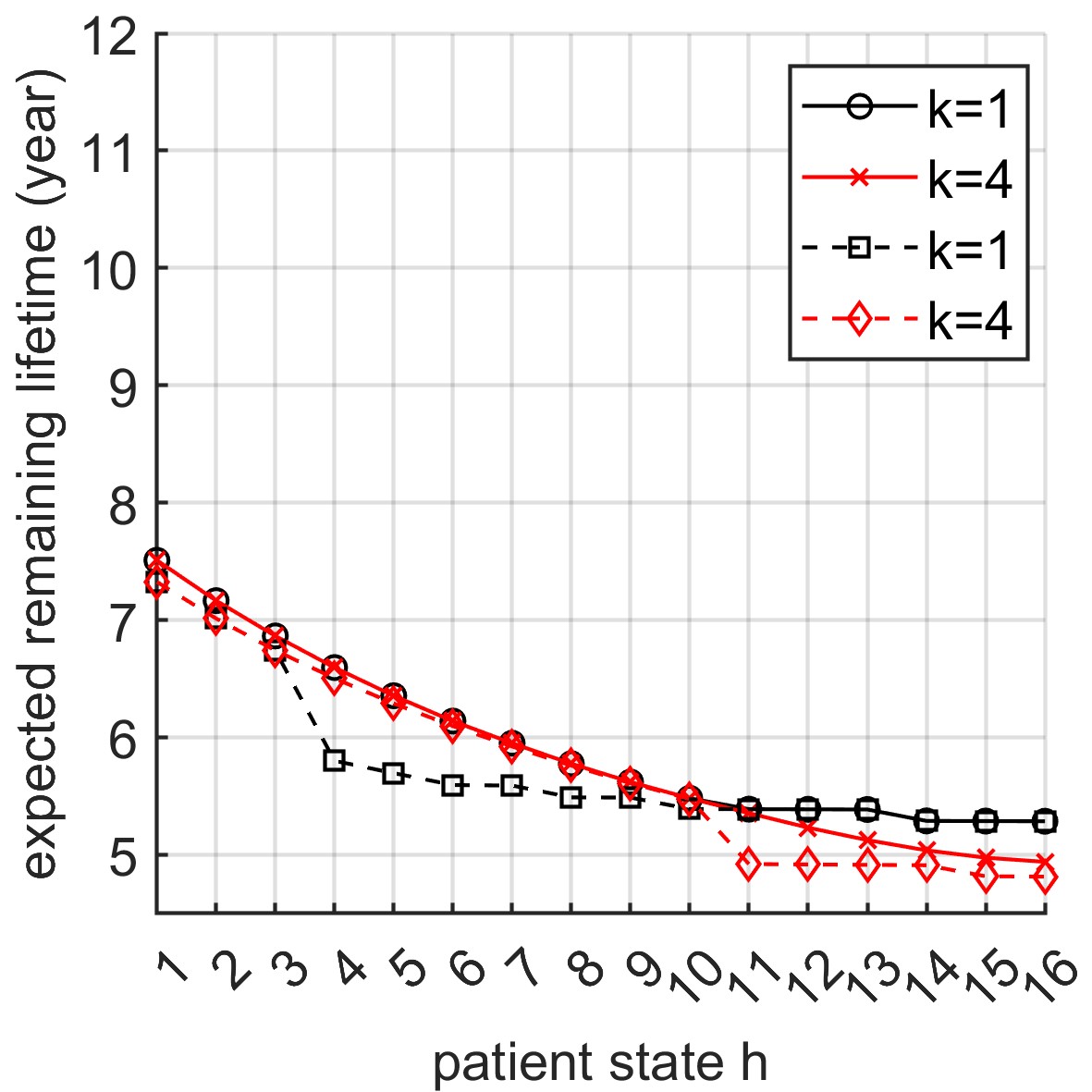}}}\hspace{5pt}
    \caption{Value functions $V_{d_1}(h,k,m)$ and $V_{\hat d_1}(h,k,m)$ for different mismatch levels, with $V_{d_1}$ represented by solid lines and $V_{\hat d_1}$ by dashed lines. For policy $\hat d_1$, fixing $m$, curves for different values of $k$ overlap with each other when $h\leq 3$ because $\hat d_1$ is always $W$ for $h\leq 3$, independent of $m$. Since both $d_1$ and $\hat d_1$ are control limit policies, fixing $k$ and $m$, if $h$ is greater than control limits of both policies, both policies choose transplant. Moreover, \Cref{patienthealth} implies that the patient state $h_t$ is a nondecreasing function of time, i.e., the patient health never improves. So, fixing $k$ and $m$, curves for both policies overlap when $h$ is greater than the maximum of control limits of both (patient-based) policies, which is a function of both $k$ and $m$.}\label{values-m}
\end{figure}

\clearpage
Next, we compare with an optimal policy derived for an MDP model that does not include the mismatch level as a state variable, i.e., the acceptance decision making depends only on the patient and kidney states, and does not explicitly model transplantation failure or retransplantation. In the case of the latter, there is a terminal reward that is the mean over the mismatch distribution. We compute its optimal policy $q_1$, shown in \Cref{2d-2-ex1}. Both patient-based and kidney-based control limit optimal policies exist. The control limit curve of policy $q_1$ lies between curves of policy $d_1$ for mismatch level $m=4$ and $5$.

To evaluate the policy $q_1(h,k)$ in the original MDP with a state vector $(h,k,m)$, we define and implement a new policy $\hat d_1(h,k,m):=q_1(h,k),~\forall h,k,m$. i.e., $\hat d_1(h,k,m)$ does not account for the mismatch level $m$ and produces the same action as $q_1(h,k)$. We compare its value function $V_{\hat d_1}(h,k,m)$ with the optimal one $V_{d_1}(h,k,m)$ in \Cref{values-m}. Compared with policy $d_1$, when $h$ is low (i.e., the patient is healthy), policy $\hat d_1$ rejects offers with low mismatch level, which would have been very beneficial to patients. As a result, policy $d_1$ significantly outperforms $\hat d_1$ when the mismatch level $m$ is low, e.g., $m\leq 3$. At intermediate mismatch levels, the two policies behave similarly. When the mismatch level is high, e.g., $m\geq 6$, the two policies behave differently only when $h$ is high (i.e., the patient is in poor health state), where the difference of the Q-functions (i.e., the expected overall benefit) between accepting and rejecting an offer is small, as shown in \Cref{q-ex1}. Thus, the performances of the two policies are close when the mismatch level is intermediate to high.



\subsection*{\hypertarget{exper2}{Experiment 2}}
\begin{figure}[H]
    \centering
    \subfloat[Given $(h,k)$, the optimal action is to wait ($W$) if and only if the mismatch level $m$ is above the surface.]{%
    \resizebox*{6.9cm}{!}{\includegraphics{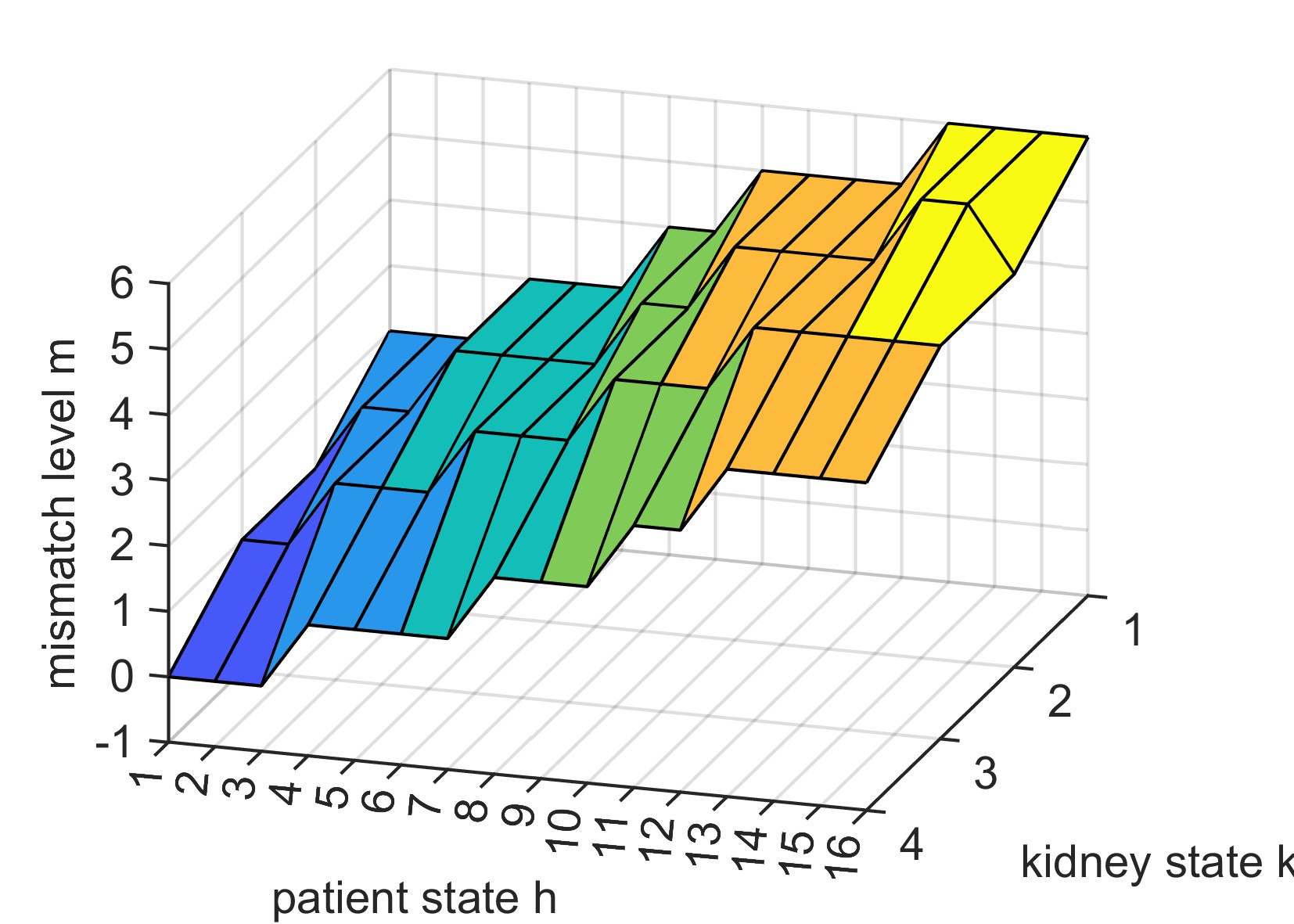}}}\hspace{5pt}
    \subfloat[Given patient state $h$, the optimal action is to wait ($W$) if and only if the kidney state $k$ is above the corresponding curve.]{%
    \resizebox*{5cm}{!}{\includegraphics{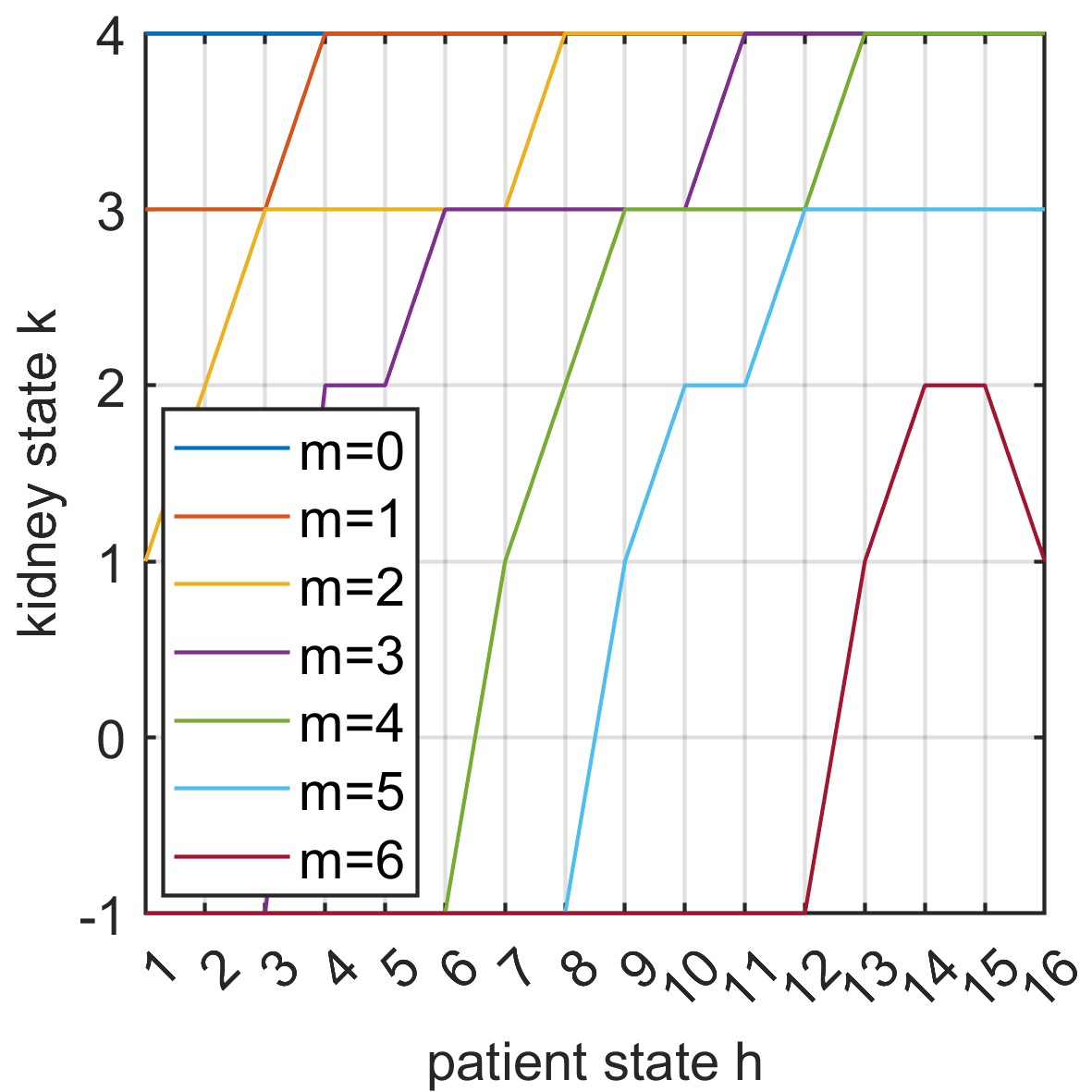}}}\hspace{5pt}
    \caption{The optimal policy $d_2$. In the graph on the right, for mismatch level $m=7$, the optimal policy is not of control limit type.} \label{3d-1-ex2}
\end{figure}

The optimal policy $d_2$ is shown in \Cref{3d-1-ex2}. The kidney-based and match-based control limit optimal policies still exist, but the patient-based optimal policy for mismatch level equal to $7$ is not a control limit policy.

\begin{figure}[ht]
    \centering
    \resizebox*{5cm}{!}{\includegraphics{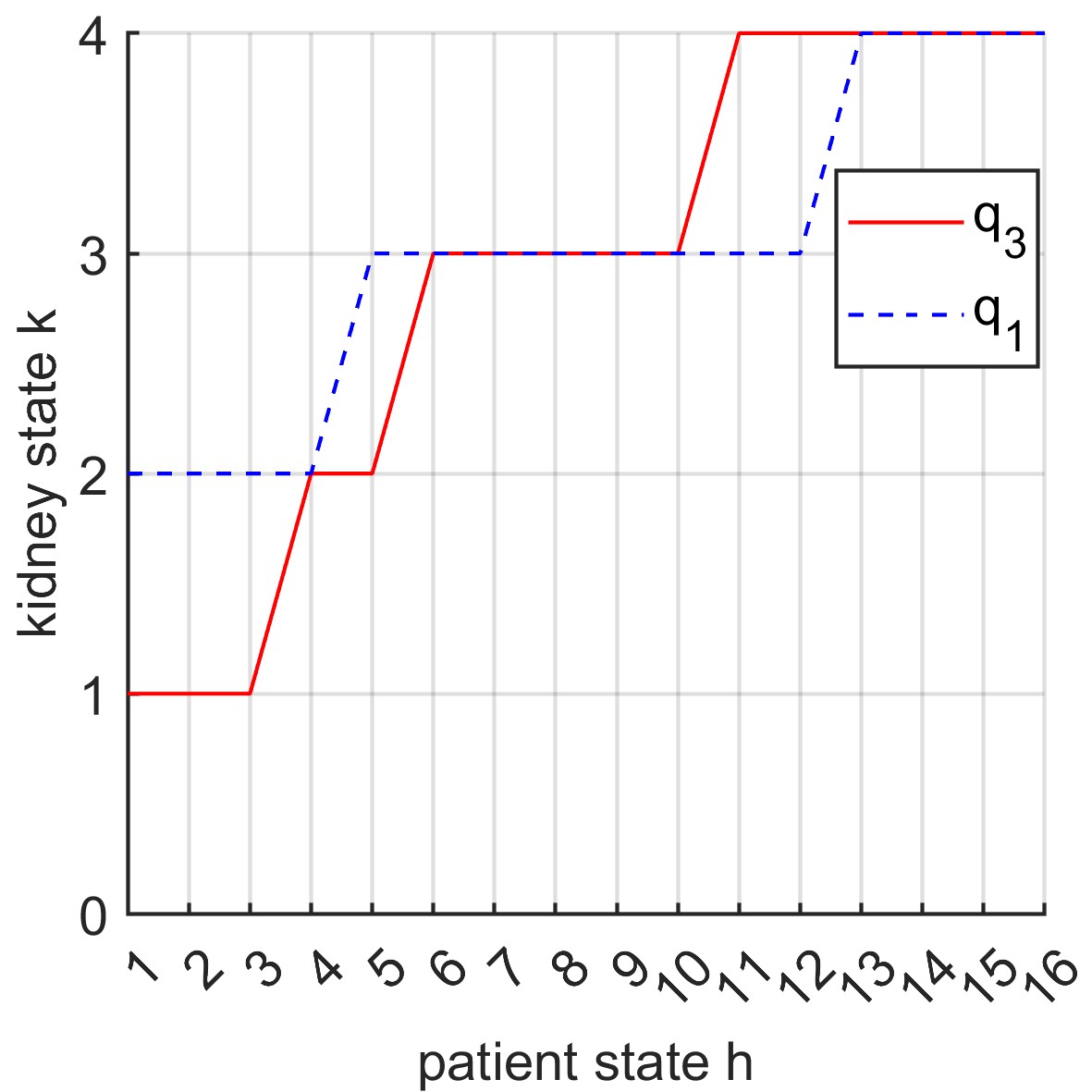}}
    \caption{Policies $q_3$ (solid line) and $q_1$ (dashed line). The control limit curve of policy $q_3$ falls below that of policy $q_1$ for $h\leq5$, and lies above it for $h=12$ and $13$.}\label{2d-1-ex3}
\end{figure}

\subsection*{\hypertarget{exper3}{Experiment 3}}
In \hyperlink{exper1}{Experiment 1}, the probability of receiving a kidney per period is below $25\%$, while the chance of transplantation failure is below $10\%$. Even if the decision maker always chooses transplantation when a kidney is available, the probability of transplantation failure remains below $1/40$ per period, indicating that retransplantation is rare. Notably, even assuming every transplantation is successful, the optimal policy would remain unchanged. Thus, in \hyperlink{exper1}{Experiment 1}, policy $d_1$ outperforms $\hat d_1$ primarily because $d_1$ accounts for the mismatch level, while the impact of retransplantation is relatively minor.

To assess the impact of retransplantation separately from the mismatch level, we consider an MDP model where transplantation failure occurs, but the mismatch level is hidden—similar to the derivation procedure of policy $q_1$ which hides both factors. The transplantation reward and the probability of transplantation failure when the mismatch level is hidden are calculated by taking the expectation over the distribution of the mismatch level. To highlight the effect of transplantation failure, we double both the probability of receiving a kidney offer and the probability of failure, leaving other parameters unchanged. We compute the optimal policy $q_3$ and compare it with policy $q_1$, as shown in \Cref{2d-1-ex3}.

The control limit curve of policy $q_3$ falls below that of policy $q_1$ for $h\leq5$, and lies above it for $h=12$ and $13$. This aligns with \Cref{h-based-con3} and the trends in terminal transplantation rewards shown in \Cref{ptreward}. Specifically, \Cref{h-based-con3} suggests that transplantation failure has a more significant impact on healthier patients, while \Cref{ptreward} indicates that the transplantation reward decline rapidly as the patient state worsens when $h\leq5$. These observations imply that the penalty for transplantation failure is more substantial for patients in better state, leading to a stricter acceptance criterion (i.e., a lower control limit curve). For $h>6$, the transplantation reward decreases only slightly as patient state deteriorates. Specifically, for patients in intermediate or worse states, even if they experience a transplantation failure and their state worsens, the decline in transplantation reward is small. Rather than imposing a penalty, explicitly modeling the transplantation failure event essentially provides patients in poor states with more opportunities for retransplantation, making them more inclined to accept a kidney.

\section{Conclusions and Future Research}\label{sec5}
We consider the problem of sequentially accepting or declining possibly incompatible kidney offers by undergoing desensitization therapies through the use of an MDP model that captures the effect of both quality and compatibility by explicitly including them as state variables. We derive structural properties of the model, specifically, characterizing control limit-type policies, including patient-based, kidney-based and match-based control limit optimal policies, by providing reasonable sufficient conditions to guarantee their existence. Numerical experiments on a stylized example based on realistic data \citep{unos2022} illustrate that the potential gains from taking both quality and compatibility into consideration and allowing the option of accepting a mismatched kidney can be on the order of an additional year of life expectancy for a 70-year old, whose expected remaining lifetime is less than five years without kidney transplantation. 


Our results have concrete applications in clinical decision-making and policy design. By explicitly accounting for patient state, kidney quality, and compatibility, our model could be integrated into decision-support tools that personalize decision-making based on patient conditions, offering simple acceptance thresholds as references for surgeons. The model's value function and Q-function quantify the gains of accepting a transplant versus waiting, providing clear measures of ``good quality'' and ``low mismatch level''. This information aids surgeons in making informed decisions and enhances transparency by enabling physicians to clearly explain trade-offs to patients, improving shared decision-making. Beyond clinical use, the acceptance criteria and quantified gains from our model could refine kidney allocation policies. These measures could enhance allocation scoring models to improve the overall transplant outcomes while balancing equity, particularly for patients with rare HLA types or blood types. The patient-specific acceptance thresholds could guide kidney match policies by estimating the likelihood of acceptance, reducing organ rejection and wastage, and ensuring more effective kidney utilization. 

Our model uses aggregated scores like EPTS and KDPI to model the state of patients and donor kidneys, making it feasible to compute the optimal policy and providing a straightforward reference tool for deciding whether to accept or decline marginal kidney offers. However, transplant surgeons consider a broader range of factors, including demographic details and additional lab results not covered by EPTS and KDPI. These factors significantly influence transplant outcomes and patient priority rankings in KAS or KPD, affecting kidney offer availability. Additionally, these features may have complex correlations that aggregated scores cannot fully capture. While an MDP model with a state vector incorporating these features would be more accurate and realistic, it would also increase the dimension of the state space, making computation intractable. Exploring the impact of these additional factors and balancing computational feasibility with model accuracy is a promising area for future research. Based on our research, it is evident that the patient's decision can significantly differ depending on the level of mismatch. Therefore, from the standpoint of policy makers, integrating our model -- which explicitly accounts for the possibility of accepting an incompatible kidney -- into the the kidney allocation problem could potentially help further reduce organ shortages and enhance overall social welfare.

\section*{Acknowledgments}
This work was supported in part by the National Science Foundation under Grant IIS-2123684 and by AFOSR under Grant FA95502010211. The authors thank Naoru Koizumi for many useful discussions on incompatible kidney transplantation.

\section*{Disclosure statement}
The authors report there are no competing interests to declare.

\bibliographystyle{apacite}
\bibliography{cas-refs}

\newpage
\appendix
\section{Proofs}\label{appenda}
Define
\begin{align}\label{sumvalue}
\begin{split}
    &U(h,k):=\sum_{m\in S_M}\M(m)V(h,k,m),~ h\in S_H,~k\in S_K,\\
    &W(h,m):=\sum_{k\in S_k} \K(k|h) V(h,k,m),~ h\in S_H,~m\in S_M,
\end{split}
\end{align}
which will be used in the proofs. Note that $U(h,k)$ and $W(h,m)$ have similar interpretations as $v(h)$, which can be interpreted as the expected total discounted reward when the patient state is $h$. Correspondingly, we define the following quantities associated with each iteration of the value iteration algorithm:
\begin{align}\label{sumvalue-iteration-dis}
\begin{split}
    &U_n(h,k)=\sum_{m\in S_M}\M(m)V_n(h,k,m),~ h\in S_H,~k\in S_K,\\
    &W_n(h,m)=\sum_{k\in S_k} \K(k|h) V_n(h,k,m),~ h\in S_H,~m\in S_M.
\end{split}
\end{align}

We begin with \Cref{puterman472-dis,puterman472col-dis}, which are useful for proving \Cref{alagoz-lemma2}, an important intermediate result for proving \Cref{mono-hk-dis}. 

\begin{deflemma} \label{puterman472-dis} \citep{puterman2014markov}
Let $\{x_j\}_{j\in\N},\{x_j'\}_{j\in\N}$ be real-valued nonnegative sequences satisfying 
\begin{align*}
    \sum_{j=k}^\infty x_j \geq \sum_{j=k}^\infty x_j'
\end{align*}
for all $k$, with equality holding for $k=0$. Suppose $v_{j+1}\geq v_j,~\forall j$, then
\begin{align*}
    \sum_{j=0}^\infty v_jx_j \geq \sum_{j=0}^\infty v_jx_j'.
\end{align*}
\end{deflemma}

\Cref{puterman472col-dis} immediately follows from \Cref{puterman472-dis}.
\begin{deflemma}\label{puterman472col-dis}
If $\H$ is stochastically increasing and $f:\R\mapsto \R$ is nondecreasing, then
\begin{align*}
    \sum_{h\in S_H} f(h) \H(h|h_1)\geq \sum_{h\in S_H} f(h) \H(h|h_2)
\end{align*}
for any $h_1\geq h_2$.
\end{deflemma}


\begin{deflemma}\label{alagoz-lemma2}
    If \Cref{transplant-reward-dis,wait-reward-dis,desenfail-dis,markovifr-dis,diffifr-dis,mono-k-assump} hold, and if $V(h,k,m)$ is nonincreasing in $h$ and $k$, then $v(h)$ is nonincreasing in $h$.
\end{deflemma}

\subsubsection*{Proof of \Cref{alagoz-lemma2}}
Recall the Bellman equation \Cref{optimality} where the second equation follows by noticing that rejecting an offer is equivalent to the offer being unavailable, and that $A^*(h,k,m)$ is the set of optimal actions at state $(h,k,m)$. 

Fix $h\leq H$ and $m\in S_M$. Define
\begin{align*}
    K_T&:=\{k\in S_K|T\in A^*(h+1,k,m)\}.
\end{align*}
Note that if $k\in K_T$, then $k\leq K$, i.e., the kidney offer is available. Then, \Cref{mono-k-assump} implies that $\K(k|h+1)\leq \K(k|h)$.

For $k\in K_T$, since $V(h,k,m)$ is nonincreasing in $h$ and $k$,
\begin{align*}
    &~~~~V(h,k,m)\K(k|h)-V(h+1,k,m)\K(k|h+1)\\&\geq V(h+1,k,m)\left(\K(k|h)-\K(k|h+1)\right)\\
    &\geq V(h+1,K+1,m)\left(\K(k|h)-\K(k|h+1)\right).
\end{align*}

For $k\notin K_T$, $V(h+1,k,m)=V(h+1,K+1,m)$. Hence,
\begin{align*}
    &~~~~V(h,k,m)\K(k|h)-V(h+1,k,m)\K(k|h+1)\\&\geq V(h,K+1,m)\K(k|h)-V(h+1,K+1,m)\K(k|h+1)\\
    &\geq  V(h+1,K+1,m)(\K(k|h)-\K(k|h+1)).
\end{align*}

Therefore,
\begin{align*}
    W(h,m)-W(h+1,m)&=\sum_{k\in K_T}(V(h,k,m)\K(k|h)-V(h+1,k,m)\K(k|h+1))\\
    &+\sum_{k\notin K_T} (V(h,k,m)\K(k|h)-V(h+1,k,m)\K(k|h+1))\\
    &\geq  \sum_{k\in S_K}(\K(k|h)-\K(k|h+1)) V(h+1,K+1,m) \\
    &= 0,
\end{align*}
because $\sum_{k\in S_K}(\K(k|h)-\K(k|h+1))= 0$. It follows that
\begin{align*}
    v(h)-v(h+1)=\sum_{m\in S_M}\M(m) (W(h,m)-W(h+1,m)) \geq 0.~ \blacksquare 
\end{align*}

\subsubsection*{Proof of \Cref{mono-hk-dis}}
Prove by induction. Consider the value iteration algorithm \Cref{value-iteration-dis} starting at $V_0(h,k,m)=0,~\forall h,k$, and $m$. Outline of the proof: at each step $n$, we show:
\begin{enumerate}
    \item $V_n(h,k,m)$ is nonincreasing in $k$. Define
    \begin{align*}
    &~~~~K_n^*(h,m) = \max\{k\in S_K \ | \ V_n(h,k,m)\\
    &=(1-\F(h,k,m))r(h,k,m)+\F(h,k,m) (c(h)+\lambda \sum_{h'\in S_H}v_{n-1}(h') \Q(h'|h)) \},
    \end{align*}
    where we assume without loss of generality (as we show below) that
    \begin{align}\label{hk-dis-emp}
        \begin{split}
            &~~~~\{k\in S_K \ | \ V_n(h,k,m)
            =(1-\F(h,k,m))r(h,k,m)+\F(h,k,m) (c(h)\\
            &+\lambda \sum_{h'\in S_H}v_n(h') \Q(h'|h))  \}\neq \phi.            
        \end{split}
    \end{align}
    We show that for fixed $h$ and $m$, \text{if } $k\leq K_n^*(h,m)$,
    \begin{align*}
        &~~~~V_n(h,k,m)\\ 
        &=(1-\F(h,k,m))r(h,k,m)+\F(h,k,m) (c(h)+\lambda \sum_{h'\in S_H}v_{n-1}(h') \Q(h'|h)),
    \end{align*}
    and $V_n(h,k,m) = c(h)+\lambda \sum_{h'\in S_H}v_{n-1}(h') \H(h'|h) \text{ if }k> K_n^*(h,m).$ Then, we show that $V_n(h,k,m)$ is nonincreasing in $k$ on both pieces. Note that the proof of $V_n(h,k,m)$ being nonincreasing is trivial if \Cref{hk-dis-emp} does not hold, because $V_n(h,k,m)=c(h)+\lambda \sum_{h'\in S_H}v(h') \H(h'|h),\\\forall k \in S_K$, which does not depend on $k$. From now on, we use the abbreviation $K_n^*$ for $K_n^*(h,m)$.
    \item $V_n(h,k,m)$ and $v_n(h)$ are nonincreasing in $h$, where we will use \Cref{alagoz-lemma2}.
\end{enumerate}

Finally, the value function $V$, as the limit of the sequence $\{V_n\}$, is nonincreasing in both $h$ and $k$. 

\textbf{Initial step:} by \Cref{value-iteration-dis}, for any $h$ and $m$, we have  
\begin{align*}
    &~~~~V_1(h,k,m)\\
    &=\begin{cases}
    \max \left( (1-\F(h,k,m))r(h,k,m)+\F(h,k,m)c(h),c(h) \right) &\text{if }k=0,\cdots,K,\\
    c(h) &\text{if }k= K+1.
    \end{cases}
\end{align*}

Then,
\begin{align*}
    &~~~~~~(1-\F(h,K_1^*,m))r(h,K_1^*,m)+\F(h,K_1^*,m)c(h)\geq c(h),\\
    &\implies (1-\F(h,K_1^*,m))(r(h,K_1^*,m) - c(h)) \geq 0,\\
    &\implies r(h,K_1^*,m) - c(h) \geq 0, \\
    &\implies r(h,k,m) - c(h) \geq 0 ,~\forall k \leq K_1^*,
\end{align*}
where the last step follows from \Cref{transplant-reward-dis} that $r(h,k,m)$ is nonincreasing in k. Therefore, $(1-\F(h,k,m))(r(h,k,m) - c(h)) \geq 0,~ \forall k \leq  K_1^*$, i.e., 
\begin{align*}
    (1-\F(h,k,m))r(h,k,m)+\F(h,k,m)c(h)\geq c(h),~k \leq K_1^*.
\end{align*}
Thus,
\begin{align*}
        V_1(h,k,m) = \begin{cases}
        (1-\F(h,k,m))r(h,k,m)+\F(h,k,m) c(h) &\text{if }k\leq K_1^*,\\
        c(h) &\text{if }k> K_1^*.
        \end{cases}
\end{align*}
For $k \leq K_1^*$, $V_1(h,k,m)=(1-\F(h,k,m))(r(h,k,m)-c(h))+c(h)$. Since both $(1-\F(h,k,m))$ and $(r(h,k,m)-c(h))$ are positive and nonincreasing in $k$ for $k \leq K_1^*$, $V_1(h,k,m)$ is nonincreasing in $k$ for $k \leq K_1^*$. Moreover, $V_1(h,k,m)\geq c(h)$ for $k \leq K_1^*$. Therefore, $V_1(h,k,m)$ is nonincreasing in $k$.

From \Cref{value-iteration-dis}, we see that $V_1(h,k,m)$ takes the value of either term in the maximization. Therefore, to show $V_1(h,k,m)\geq V_1(h+1,k,m)$ for any $k$ and $m$, it suffices to consider the following four cases:
\begin{enumerate}
    \item $V_1(h,k,m) =  c(h),~V_1(h+1,k,m) = c(h+1)$. Proving $V_1(h,k,m)\geq V_1(h+1,k,m)$ is trivial, as $c(h)\geq c(h+1)$ by \Cref{wait-reward-dis}.
    \item $V_1(h,k,m) =  (1-\F(h,k,m))r(h,k,m)+\F(h,k,m) c(h),~V_1(h+1,k,m) = c(h+1)$. Then,
    \begin{align*}
        V_1(h,k,m) &=  (1-\F(h,k,m))r(h,k,m)+\F(h,k,m) c(h) \geq c(h) \geq c(h+1) \\
        &= V_1(h+1,k,m).
    \end{align*}
    \item $V_1(h,k,m) =  (1-\F(h,k,m))r(h,k,m)+\F(h,k,m) c(h),~V_1(h+1,k,m) =(1-\\ \F(h+1,k,m))r(h+1,k,m)+\F(h+1,k,m) c(h+1)$. Note that 
    \begin{align*}
        (1- \F(h,k,m))r(h,k,m)+\F(h,k,m) c(h) \geq c(h).
    \end{align*}
    
    Then,
    \begin{align*}
        (1- \F(h,k,m))(r(h,k,m)-c(h)) \geq 0\implies r(h,k,m)-c(h)\geq 0.
    \end{align*}
    
    Therefore,
    \begin{align}\label{proof-mono-hk-1}
    \begin{split}
        &~~~~V_1(h,k,m) - V_1(h+1,k,m)\\
        & = (1-\F(h,k,m))r(h,k,m)+\F(h,k,m) c(h) - (1- \F(h+1,k,m))r(h+1,k,m)\\ 
        &- \F(h+1,k,m) c(h+1)\\
        &=(1-\F(h,k,m))r(h,k,m)+\F(h,k,m) c(h) - (1- \F(h+1,k,m))r(h+1,k,m) \\
        &- \F(h,k,m) c(h+1) + (\F(h,k,m) - \F(h+1,k,m)) c(h+1)\\
        &\geq (1-\F(h,k,m))r(h,k,m)+\F(h,k,m) c(h) - (1- \F(h+1,k,m))r(h+1,k,m) \\
        &- \F(h,k,m) c(h+1) + (\F(h,k,m) - \F(h+1,k,m)) r(h,k,m)\\
        &= (1-\F(h+1,k,m))(r(h,k,m)-r(h+1,k,m))+\F(h,k,m) (c(h)-c(h+1))\\
        &\geq 0,
    \end{split}
    \end{align}
    where the first inequality follows from the fact that $\F(h,k,m) \leq \F(h+1,k,m)$ and $r(h,k,m)\geq c(h) \geq c(h+1)$, and the last inequality follows from \Cref{wait-reward-dis,transplant-reward-dis}.
    \item $V_1(h,k,m) =  c(h),~V_1(h+1,k,m) = (1- \F(h+1,k,m))r(h+1,k,m)+\F(h+1,k,m) c(h+1)$. Note that 
    \begin{align*}
        (1- \F(h+1,k,m))r(h+1,k,m)+\F(h+1,k,m) c(h+1) \geq c(h+1).
    \end{align*}
    
    Then,
    \begin{align*}
        (1- \F(h+1,k,m))(r(h+1,k,m)-c(h+1)) \geq 0&\implies r(h+1,k,m)-c(h+1)\geq 0 \\ &\implies c(h+1) \leq r(h,k,m).
    \end{align*}
    Notice that $V_1(h,k,m)\geq (1-\F(h,k,m))r(h,k,m)+\F(h,k,m) c(h)$. Using the same argument as \Cref{proof-mono-hk-1}, we have $V_1(h,k,m) \geq V_1(h+1,k,m)$.
\end{enumerate}
 Thus, $V_1(h,k,m)$ is nonincreasing in $h$. \Cref{alagoz-lemma2} implies that $v_1(h)$ is nonincreasing in $h$.

\textbf{Induction step:} suppose that $V_n(h,k,m)$ is nonincreasing in both $h$ and $k$, and $v_n(h)$ is nonincreasing in $h$. By \Cref{value-iteration-dis},
\begin{align*}
\begin{split}
&~~~~ V_{n+1}(h,k,m)\\
&=\max\begin{pmatrix}
    (1-\F(h,k,m))r(h,k,m)+\F(h,k,m) (c(h)+\lambda \sum_{h'\in S_H}v_n(h') \Q(h'|h))\\
    c(h)+\lambda \sum_{h'\in S_H}v_n(h') \H(h'|h)
    \end{pmatrix},\\
&V_{n+1}(h,K+1,m)= c(h)+\lambda \sum_{h'\in S_H}v_n(h') \H(h'|h),
\end{split}
\end{align*}
$\\~\forall h\in S_H,~m \in S_M,~k<K+1$. Then, for any $h$ and $m$,
\begin{align*}
    &~~~~(1-\F(h,K_n^*,m))r(h,K_n^*,m)+\F(h,K_n^*,m) (c(h)+\lambda \sum_{h'\in S_H}v_n(h') \Q(h'|h))\\
    &=(1-\F(h,K_n^*,m))\left(r(h,K_n^*,m)- c(h)-\lambda \sum_{h'\in S_H}v_n(h') \Q(h'|h)  \right)\\
    & + c(h)+\lambda \sum_{h'\in S_H}v_n(h') \Q(h'|h)\\
    &\geq c(h)+\lambda \sum_{h'\in S_H}v_n(h') \H(h'|h).
\end{align*}
It follows that
\begin{align*}
    &~~~~(1-\F(h,K_n^*,m))\left(r(h,K_n^*,m)- c(h)-\lambda \sum_{h'\in S_H}v_n(h') \Q(h'|h)  \right)\\ &\geq \lambda \sum_{h'\in S_H}v_n(h') (\H(h'|h) - \Q(h'|h))\\ &\geq 0,
\end{align*}
where the last inequality follows from \Cref{diffifr-dis} that $\Q\succeq_{st} \H$ and \Cref{puterman472col-dis}. Since $r(h,k,m)$ is nonincreasing in $k$,
\begin{align*}
    r(h,k,m)- c(h)-\lambda \sum_{h'\in S_H}v_n(h') \Q(h'|h) \geq 0,~\forall k \leq K_n^*.
\end{align*}
Since $\F(h,k,m)$ is nondecreasing in $k$, $\forall k \leq  K_n^*$,
\begin{align*}
    &~~~~(1-\F(h,k,m))\left(r(h,k,m)- c(h)-\lambda \sum_{h'\in S_H}v_n(h') \Q(h'|h)  \right)\\& \geq \lambda \sum_{h'\in S_H}v_n(h') (\H(h'|h) - \Q(h'|h)),
\end{align*}
i.e., 
\begin{align*}
    &~~~~(1-\F(h,k,m))r(h,k,m)+\F(h,k,m) (c(h)+\lambda \sum_{h'\in S_H}v_n(h') \Q(h'|h))\\&\geq 
    c(h)+\lambda \sum_{h'\in S_H}v_n(h') \H(h'|h).
\end{align*}
Therefore,
\begin{align*}
\begin{cases}
V_{n+1}(h,k,m)=
    (1-\F(h,k,m))r(h,k,m)\\+\F(h,k,m) (c(h)+\lambda \sum_{h'\in S_H}v_n(h') \Q(h'|h)) &\text{if } k\leq K_n^*,\\
V_{n+1}(h,k,m)= c(h)+\lambda \sum_{h'\in S_H}v_n(h') \H(h'|h) &\text{if } k> K_n^*.
\end{cases}
\end{align*}
Moreover, when $k\leq K_n^*$,
\begin{align*}
    &~~~~(1-\F(h,k,m))r(h,k,m)+\F(h,k,m) (c(h)+\lambda \sum_{h'\in S_H}v_n(h') \Q(h'|h))\\
    &=(1-\F(h,k,m))\left(r(h,k,m)- c(h)-\lambda \sum_{h'\in S_H}v_n(h') \Q(h'|h)  \right)\\ &+ c(h)+\lambda \sum_{h'\in S_H}v_n(h') \Q(h'|h),
\end{align*}
where both $(1-\F(h,k,m))$ and $\left(r(h,k,m)- c(h)-\lambda \sum_{h'\in S_H}v_n(h') \Q(h'|h)  \right)$ are positive and nonincreasing in $k$, so $(1-\F(h,k,m))r(h,k,m)+\F(h,k,m) (c(h)+\lambda \sum_{h'\in S_H}v_n(h') \Q(h'|h))$ is nonincreasing in $k$ for $k\leq K_n^*$. Moreover, $(1-\F(h,k,m))r(h,k,m)+\F(h,k,m) (c(h)+\lambda \sum_{h'\in S_H}v_n(h') \\ \times\Q(h'|h))\geq c(h)+\lambda \sum_{h'\in S_H}v_n(h') \H(h'|h)$ for $k\leq K_n^*$. Therefore, $V_{n+1}(h,k,m)$ is nonincreasing in $k$.

To show $V_{n+1}(h,k,m)\geq V_{n+1}(h+1,k,m)$ for any $k$ and $m$, as above, it suffices to consider the following four cases:
\begin{enumerate}
    \item $V_{n+1}(h,k,m) =  c(h)+\lambda \sum_{h'\in S_H}v_n(h') \H(h'|h),~V_{n+1}(h+1,k,m) = c(h+1)+\lambda \sum_{h'\in S_H}v_n(h') \\ \times\H(h'|h+1)$. By \Cref{wait-reward-dis}, $c(h)\geq c(h+1)$. By \Cref{markovifr-dis}, \Cref{puterman472col-dis} and $v_n(h)$ being nonincreasing, $\sum_{h'\in S_H}v_n(h') \H(h'|h)\geq  \sum_{h'\in S_H}v_n(h') \H(h'|h+1)$. So, $V_{n+1}(h,k,m)\geq V_{n+1}(h+1,k,m)$.
    \item $V_{n+1}(h,k,m) = (1-\F(h,k,m))r(h,k,m)+\F(h,k,m) (c(h)+\lambda \sum_{h'\in S_H}v_n(h') \Q(h'|h)),\\~V_{n+1}(h+1,k,m) = c(h+1)+\lambda \sum_{h'\in S_H}v_n(h') \H(h'|h+1))$. Then,
    \begin{align*}
        V_{n+1}(h,k,m) &\geq c(h)+\lambda \sum_{h'\in S_H}v_n(h') \H(h'|h)\\ &\geq c(h+1) +\lambda \sum_{h'\in S_H}v_n(h') \H(h'|h+1)= V_{n+1}(h+1,k,m),
    \end{align*}
    where the second inequality follows from the same argument as the first case.
    \item  $V_{n+1}(h,k,m) = (1-\F(h,k,m))r(h,k,m)+\F(h,k,m) (c(h)+\lambda \sum_{h'\in S_H}v_n(h') \Q(h'|h)),\\~V_{n+1}(h+1,k,m) = (1-\F(h+1,k,m))r(h+1,k,m)+\F(h+1,k,m) (c(h+1)+\lambda \sum_{h'\in S_H}v_n(h')  \\ \times \Q(h'|h+1))$. Note that $\E g((h,k,m),T) =(1-\F(h,k,m))r(h,k,m)+\F(h,k,m) c(h),~\forall h,k,$ and $m$. Then,
    \begin{align}\label{proof-mono-hk-2}
        \begin{split}
        &~~~~ V_{n+1}(h,k,m)-V_{n+1}(h+1,k,m)\\
        &= \E g((h,k,m),T)+\F(h,k,m)\lambda \sum_{h'\in S_H}v_n(h') \Q(h'|h) \\
        &- \E g((h+1,k,m),T) - \F(h+1,k,m)\lambda \sum_{h'\in S_H}v_n(h') \Q(h'|h+1)\\
        &=\E g((h,k,m),T) - \E g((h+1,k,m),T)\\ 
        &+ (\F(h,k,m) - \F(h+1,k,m))\lambda \sum_{h'\in S_H}v_n(h') \Q(h'|h)\\
        &+ \F(h+1,k,m)\lambda \sum_{h'\in S_H}v_n(h')( \Q(h'|h) - \Q(h'|h+1))\\
        &\geq \E g((h,k,m),T) - \E g((h+1,k,m),T)\\ 
        &+ (\F(h,k,m) - \F(h+1,k,m))\lambda \sum_{h'\in S_H}v_n(h') \Q(h'|h),
        \end{split}
    \end{align}
    where the last inequality follows from \Cref{puterman472col-dis}, \Cref{markovifr-dis} and $v_n(h)$ being nonincreasing. Note that $V_{n+1}(h,k,m) \geq c(h)+\lambda \sum_{h'\in S_H}v_n(h') \H(h'|h)$, i.e.,
    \begin{align*}
        &~~~~(1-\F(h,k,m))\left (r(h,k,m) - c(h) - \lambda \sum_{h'\in S_H}v_n(h') \Q(h'|h) \right)\\
        &\geq  \lambda \sum_{h'\in S_H}v_n(h')(\H(h'|h) - \Q(h'|h))\\
        &\geq 0,
    \end{align*}
    where the last inequality follows from \Cref{puterman472col-dis} and \Cref{diffifr-dis}. So,
    \begin{align*}
        r(h,k,m) - c(h) \geq  \lambda \sum_{h'\in S_H}v_n(h') \Q(h'|h).
    \end{align*}
    Since $\F(h,k,m) - \F(h+1,k,m)\leq 0$, by \Cref{proof-mono-hk-2},
    \begin{align*}
        &~~~~ V_{n+1}(h,k,m)-V_{n+1}(h+1,k,m)\\
        &\geq \E g((h,k,m),T) - \E g((h+1,k,m),T)\\ 
        &+ (\F(h,k,m) - \F(h+1,k,m))(r(h,k,m) - c(h))\\
        &= (1-\F(h,k,m))r(h,k,m)+\F(h,k,m) c(h)\\ 
        &- (1-\F(h+1,k,m))r(h+1,k,m)- \F(h+1,k,m) c(h+1) \\
        &+ (\F(h,k,m) - \F(h+1,k,m))(r(h,k,m) - c(h))\\
        &=(1-\F(h+1,k,m))(r(h,k,m) - r(h+1,k,m))\\ 
        &+ \F(h+1,k,m) (c(h)-c(h+1))\\
        &\geq 0,
    \end{align*}
    where the last inequality follows from \Cref{wait-reward-dis,transplant-reward-dis}. 
    \item $V_{n+1}(h,k,m) =   c(h)+\lambda \sum_{h'\in S_H}v_n(h') \H(h'|h),~V_{n+1}(h+1,k,m) = (1-\F(h+1,k,m)) \\ \times r(h+1,k,m)+\F(h+1,k,m) (c(h+1)+\lambda \sum_{h'\in S_H}v_n(h') \Q(h'|h+1))$. Note that 
    \begin{align*}
        &~~~~(1-\F(h+1,k,m))r(h+1,k,m)+\F(h+1,k,m) (c(h+1)\\&+\lambda \sum_{h'\in S_H}v_n(h') \Q(h'|h+1)) \\&\geq c(h+1)+\lambda \sum_{h'\in S_H}v_n(h') \H(h'|h+1)).
    \end{align*}
    Then,
    \begin{align*}
        &~~~~(1-\F(h+1,k,m))\left (r(h+1,k,m) - c(h+1) - \lambda \sum_{h'\in S_H}v_n(h') \Q(h'|h+1) \right)\\
        &\geq  \lambda \sum_{h'\in S_H}v_n(h')(\H(h'|h+1) - \Q(h'|h+1))\\
        &\geq 0,
    \end{align*}
    where the last inequality follows from \Cref{puterman472col-dis}, \Cref{diffifr-dis} and $v_n$ being nonincreasing. So,
    \begin{align}\label{proof-mono-hk-3}
          c(h+1) + \lambda \sum_{h'\in S_H}v_n(h') \Q(h'|h+1) \leq r(h+1,k,m).
    \end{align}
    Then,
    \begin{align*}
        &~~~~ V_{n+1}(h,k,m)-V_{n+1}(h+1,k,m)\\
        &= \E g((h,k,m),T)+\F(h,k,m)\lambda \sum_{h'\in S_H}v_n(h') \Q(h'|h) \\
        &- \E g((h+1,k,m),T) - \F(h+1,k,m)\lambda \sum_{h'\in S_H}v_n(h') \Q(h'|h+1)\\
        &=\E g((h,k,m),T) - \E g((h+1,k,m),T)\\ &+ \F(h,k,m) \lambda \sum_{h'\in S_H}v_n(h') (\Q(h'|h)-\Q(h'|h+1))\\
        &-(\F(h+1,k,m) -\F(h,k,m) )\lambda \sum_{h'\in S_H}v_n(h')\Q(h'|h+1)\\
        &\geq \E g((h,k,m),T) - \E g((h+1,k,m),T)\\
        &-(\F(h+1,k,m) -\F(h,k,m) )\lambda \sum_{h'\in S_H}v_n(h')\Q(h'|h+1)\\
        &\geq \E g((h,k,m),T) - \E g((h+1,k,m),T)\\ 
        &+ (\F(h,k,m) - \F(h+1,k,m)) )(r(h+1,k,m) - c(h+1))\\
        &= (1-\F(h,k,m))r(h,k,m)+\F(h,k,m) c(h)\\ 
        &- (1-\F(h+1,k,m))r(h+1,k,m)-\F(h+1,k,m) c(h+1)\\
        &+ (\F(h,k,m) - \F(h+1,k,m)) (r(h+1,k,m) - c(h+1))\\
        &= (1-\F(h,k,m))(r(h,k,m) - r(h+1,k,m) )+\F(h,k,m)( c(h) - c(h+1))\\
        &\geq 0,
    \end{align*}
    where the first inequality follows from \Cref{puterman472col-dis}, \Cref{diffifr-dis} and $v_n$ being nonincreasing, and the second inequality follows from \Cref{proof-mono-hk-3}.
\end{enumerate}

Therefore, $V_{n+1}(h,k,m)$ is nonincreasing in $h$. \Cref{alagoz-lemma2} implies that $v_{n+1}$ is nonincreasing in $h$. Finally, function $V$, as the limit of sequence $\{V_n\}$, is nonincreasing in both $h$ and $k$, and \Cref{alagoz-lemma2} implies that $v$ is nonincreasing in $h$. $\blacksquare$

\subsubsection*{Proof of \Cref{match-based-dis,kidney-based-dis}}
To show the existence of the match-based control limit policy, it is equivalent to show that $T\in A^*(h,k,m+1)$ implies $T\in A^*(h,k,m)$ for any $h,k,m$, where $A^*(h,k,m)$ is the set of optimal actions at state $(h,k,m)$. Assume that $T\in A^*(h,k,m+1)$, then
\begin{align*}
    &~~~~V(h,k,m+1)\\&=(1-\F(h,k,m+1))r(h,k,m+1)+\F(h,k,m+1)(c(h)+\lambda \sum_{h'\in S_H}v(h') \Q(h'|h))\\
    &\geq c(h)+\lambda \sum_{h'\in S_H}v(h') \H(h'|h),
\end{align*}
i.e.,
\begin{align*}
    &~~~~(1-\F(h,k,m+1)(r(h,k,m+1)-c(h))\\
    &\geq (1-\F(h,k,m+1))\lambda \sum_{h'\in S_H}v(h') \H(h'|h))\\
    &+\F(h,k,m+1)\lambda \sum_{h'\in S_H}v(h') (\H(h'|h)-\Q(h'|h)),
\end{align*}
i.e.,
\begin{align*}
    &~~~~(1-\F(h,k,m+1)(r(h,k,m+1)-c(h)-\lambda \sum_{h'\in S_H}v(h') \H(h'|h))\\
    &\geq \F(h,k,m+1)\lambda \sum_{h'\in S_H}v(h') (\H(h'|h)-\Q(h'|h))\\
    &\geq 0,
\end{align*}
where the last inequality follows from \Cref{puterman472col-dis}, \Cref{diffifr-dis} and $v$ being nonincreasing. So, if $T\in A^*(h,k,m+1)$,
\begin{align}\label{proof-cl-mk-1}
    r(h,k,m+1)-c(h)-\lambda \sum_{h'\in S_H}v(h') \H(h'|h)\geq 0.
\end{align}

Since $\F(h,k,m)$ is nondecreasing in $m$ and $r(h,k,m)$ is nonincreasing in $m$, we have 
\begin{align*}
    &~~~~(1-\F(h,k,m)(r(h,k,m)-c(h)-\lambda \sum_{h'\in S_H}v(h') \H(h'|h))\\
    &\geq \F(h,k,m)\lambda \sum_{h'\in S_H}v(h') (\H(h'|h)-\Q(h'|h)),
\end{align*}
i.e.,
\begin{align*}
    V(h,k,m)&=(1-\F(h,k,m))r(h,k,m)+\F(h,k,m)(c(h)+\lambda \sum_{h'\in S_H}v(h') \Q(h'|h))\\
    &\geq c(h)+\lambda \sum_{h'\in S_H}v(h') \H(h'|h).
\end{align*}

Therefore, $T\in A^*(h,k,m)$. The proof of \Cref{kidney-based-dis} is exactly the same (by showing $T\in A^*(h,k+1,m)\implies T\in A^*(h,k,m)$) and is omitted. $\blacksquare$

\subsubsection*{Proof of \Cref{mono-m-dis}}
If $m\geq M^*(h,k)$, it is optimal to choose $W$, and $V(h,k,m)=c(h)+\lambda \sum_{h'\in S_H}v(h') \H(h'|h)$, as a function of $m$, is constant.

If $m< M^*(h,k)$, $T\in A^*(h,k,m)$ and
\begin{align*}
    V(h,k,m)=&(1-\F(h,k,m))r(h,k,m)+\F(h,k,m)(c(h)+\lambda \sum_{h'\in S_H}v(h') \Q(h'|h))\\
    &\geq c(h)+\lambda \sum_{h'\in S_H}v(h') \H(h'|h).
\end{align*}
In particular, $V(h,k,M^*(h,k)-1)\geq V(h,k,M^*(h,k))$. From \Cref{proof-cl-mk-1}, we know that $r(h,k,m)-c(h)-\lambda \sum_{h'\in S_H}v(h') \H(h'|h)\geq 0 $ for $m< M^*(h,k)$.

Since $\sum_{h'\in S_H}v(h') \H(h'|h) \geq \sum_{h'\in S_H}v(h') \Q(h'|h)$ (follows from \Cref{puterman472col-dis}, \Cref{diffifr-dis} and $v$ being nonincreasing), we have $r(h,k,m)-c(h)-\lambda \sum_{h'\in S_H}v(h') \Q(h'|h)\geq 0 $ for $m< M^*(h,k)$.

Rewrite
\begin{align*}
    V(h,k,m)&=(1-\F(h,k,m))\left(r(h,k,m)-c(h)-\lambda \sum_{h'\in S_H}v(h') \Q(h'|h)\right)\\&+\left(c(h)+\lambda \sum_{h'\in S_H}v(h') \Q(h'|h)\right).
\end{align*}

Since $\F(h,k,m)$ is nondecreasing in $m$ and $r(h,k,m)$ is nonincreasing in $m$, $V(h,k,m)$ is nonincreasing in $m$ for $m< M^*(h,k)$. Thus, $V(h,k,m)$ is nonincreasing in $m$. $\blacksquare$

\subsubsection*{Proof of \Cref{mono-limit-1-dis}}
When proving \Cref{match-based-dis}, we showed that for any $k,h$, $T\in a(h,k,m+1)$ implies that $T\in a(h,k,m)$, i,e, $k<K^*(h,m+1)$ implies that $k<K^*(h,m)$. Hence, $K^*(h,m+1)\leq K^*(h,m)$. Similarly, we can argue that $M^*(h,k+1)\leq M^*(h,k)$. The last two statements follow from monotonicity. $\blacksquare$

\Cref{schaefer-lemma3} is provided in \cite{alagoz2007determining}. 
\begin{deflemma}\label{schaefer-lemma3}
Suppose transition probability function $P(\cdot|\cdot):S\times S\mapsto [0,1]$ on state space $S=\{1,\cdots,n\}$ is stochastically increasing. If function $f:\R\mapsto \R_{+}$ is nonincreasing, the following inequalities hold: for $i=1,\cdots,n,~j=1,\cdots,n-1$,
\begin{enumerate}
    \item $\sum_{i\leq j}(P(i|j)-P(i|j+1))f(i) \geq \sum_{i\leq j}(P(i|j)-P(i|j+1))f(j)$;
    \item $\sum_{i> j}(P(i|j)-P(i|j+1))f(i) \geq \sum_{i> j}(P(i|j)-P(i|j+1))f(j+1)$.
\end{enumerate}
\end{deflemma}

\subsubsection*{Proof of \Cref{health-based-dis}}
It is equivalent to show that $T\in A^*(h,k,m)$ implies that $T\in A^*(h+1,k,m)$ for any $h,k,$ and $m$. Fix $h\leq H,~k\leq K$. Prove by contradiction: suppose that $T\in A^*(h,k,m)$ but $A^*(h+1,k,m)=W$. Since $T\in A^*(h,m,k)$, we have 
\begin{align}\label{proof-mono-dis-1}
\begin{split}
    V(h,k,m)&=\E g((h,k,m),T)+\F(h,k,m)\lambda \sum_{h'\in S_H}v(h') \Q(h'|h)\\
    &\geq c(h)+\lambda \sum_{h'\in S_H}v(h') \H(h'|h).
\end{split}
\end{align}
Since $A^*(h+1,m,k)=W$, the following strict inequality holds:
\begin{align}\label{proof-mono-dis-2}
\begin{split}
     &~~~~\E g((h+1,k,m),T)+\F(h+1,k,m)\lambda \sum_{h'\in S_H}v(h') \Q(h'|h+1)\\&< c(h+1)+\lambda \sum_{h'\in S_H}v(h')\H(h'|h+1)    .
\end{split}
\end{align}
Subtracting \Cref{proof-mono-dis-2} from \Cref{proof-mono-dis-1},
\begin{align*}
    &~~~~\E g((h,k,m),T)+\F(h,k,m)\lambda \sum_{h'\in S_H}v(h') \Q(h'|h) \\
    &- \E g((h+1,k,m),T) - \F(h+1,k,m)\lambda \sum_{h'\in S_H}v(h') \Q(h'|h+1)\\
    &> c(h) - c(h+1) +\lambda \sum_{h'\in S_H}v(h') \H(h'|h) - \lambda \sum_{h'\in S_H}v(h')\H(h'|h+1).
\end{align*}
Since $\F(h+1,k,m) \geq \F(h,k,m)$ (\Cref{desenfail-dis}),
\begin{align}
    \begin{split}\label{proof-mono-dis-3}
    &~~~~\E g((h,k,m),T)- \E g((h+1,k,m),T)\\&> c(h)-c(h+1) + \F(h,k,m)\lambda \sum_{h'\in S_H}v(h') (\Q(h'|h+1)-\Q(h'|h))\\
    &+\lambda \sum_{h'\in S_H}v(h') \left(\H(h'|h)-\H(h'|h+1)\right)\\
    &\geq \F(h,k,m)\lambda \sum_{h'\in S_H}v(h') (\Q(h'|h+1)-\Q(h'|h))\\
    &+\lambda \sum_{h'\in S_H}v(h') \left(\H(h'|h)-\H(h'|h+1)\right)\\
    &=\F(h,k,m)\lambda \sum_{h'\in S_H}v(h') (\Q(h'|h+1)-\H(h'|h+1)-(\Q(h'|h)-\H(h'|h)))\\
    &+\lambda (1-\F(h,k,m)) \sum_{h'\in S_H}v(h') \left(\H(h'|h)-\H(h'|h+1)\right)\\
    &\geq \lambda (1-\F(h,k,m)) \sum_{h'\in S_H}v(h') \left(\H(h'|h)-\H(h'|h+1)\right),
    \end{split}
\end{align}
where the second inequality follows from \Cref{wait-reward-dis} that $c(h)\geq c(h+1)$; to prove the last inequality, it is enough to notice that
\begin{align*}
    &~~~~\sum_{h'\in S_H}v(h') (\Q(h'|h+1)-\H(h'|h+1)-(\Q(h'|h)-\H(h'|h)))\\
    &= \sum_{h'\in S_H}v(h')(\Q(h'|h+1)+\H(h'|h)-(\H(h'|h+1)+\Q(h'|h)))\\
    &\geq 0
\end{align*}
because of \Cref{puterman472-dis}, \Cref{h-based-con3} and $v$ being nonincreasing.

From \Cref{proof-mono-dis-3}, it follows that
\begin{align}
\begin{split}\label{proof-mono-dis-4}
    &~~~~\E g((h,k,m),T)- \E g((h+1,k,m),T)\\ &>  \lambda (1-\F(h,k,m)) \sum_{h'\in S_H}v(h') \left(\H(h'|h)-\H(h'|h+1)\right)\\
    &=\lambda (1-\F(h,k,m)) \biggl( \sum_{h'=1}^h  v(h') \left(\H(h'|h)-\H(h'|h+1)\right)\\ 
    &+\sum_{h'=h+1}^H v(h') \left(\H(h'|h)-\H(h'|h+1)\right) \biggr)
\end{split}
\end{align}
(note that $v(H+1)=0$). Since $v$ is nonincreasing, by applying both parts of \Cref{schaefer-lemma3} to each of the sums in \Cref{proof-mono-dis-4}, respectively, we have
\begin{align*}
    &~~~~\E g((h,k,m),T)- \E g((h+1,k,m),T)\\ &> \lambda (1-\F(h,k,m))\biggl( v(h)\sum_{h'=1}^h \left(\H(h'|h)-\H(h'|h+1)\right)\\ 
    &+v(h+1)\sum_{h'=h+1}^H  \left(\H(h'|h)-\H(h'|h+1)\right) \biggr)\\
    &=\lambda(1-\F(h,k,m)) v(h)\\
    &\times\left(1-\sum_{h'=h+1}^H \H(h'|h) -\H(H+1|h) -\left(1-\sum_{h'=h+1}^H \H(h'|h+1) -\H(H+1|h+1)\right) \right) \\
    &+\lambda(1-\F(h,k,m)) v(h+1)\sum_{h'=h+1}^H  \left(\H(h'|h)-\H(h'|h+1)\right)\\
    &=\lambda(1-\F(h,k,m))(v(h)-v(h+1))\sum_{h'=h+1}^H   \left(\H(h'|h+1)-\H(h'|h)\right) \\
    &+\lambda(1-\F(h,k,m)) v(h) (\H(H+1|h+1)-\H(H+1|h)).
\end{align*}

By \Cref{h-based-con1}, $\sum_{h'=h+1}^H   \left(\H(h'|h+1)-\H(h'|h)\right)\geq 0$.
Therefore,
\begin{align}\label{proof-mono-dis-5}
\begin{split}
    &~~~~\E g((h,k,m),T)- \E g((h+1,k,m),T) \\
    &>  \lambda (1-\F(h,k,m)) v(h)  \left(\H(H+1|h+1)-\H(H+1|h)\right)   
\end{split}
\end{align}
From \Cref{h-based-con2} and \Cref{proof-mono-dis-5}, we have
\begin{align*}
    v(h)<\E g((h+1,k,m),T).
\end{align*}

Since $\E g((h+1,k,m),T)\leq  V(h+1,k,m)$, we obtain $v(h) < V(h+1,k,m)$. By \Cref{proof-mono-dis-2},\\ $V(h+1,k,m)=c(h+1)+\lambda \sum_{h'\in S_H}v_n(h')\H(h'|h+1)=V(h+1,K+1,m)$. Then, by the monotonicity of $V(h,k,m)$ in both $h$ and $k$,
\begin{align*}
    v(h) < V(h+1,k,m) = V(h+1,K+1,m) \leq V(h,K+1,m) \leq v(h),
\end{align*}
where the last inequality follows from the fact that $V(h,K+1,m)\leq V(h,k,m),~\forall k,m$, which is a contradiction. Therefore, $T\in A^*(h,k,m)$ implies that $T\in A^*(h+1,k,m)$. $\blacksquare$

\subsubsection*{Proof of \Cref{diff-avai}}
Prove by induction. Suppose that we solve $\Pi_1$ and $\Pi_2$ simultaneously using value iteration \Cref{value-iteration-dis} with $V_0^1(h,k,m)=V_0^2(h,k,m)=0,~\forall h,k,$ and $m$. Let $V^i_j$ be the value function of $\Pi_i,~i=1,2$ at the $j^{th}$ iteration.

\textbf{Initial step:} at the first iteration,
\begin{align*}
    V_1^1(h,k,m)&=V_1^2(h,k,m)\\
    &=\begin{cases}
    \max\left( (1-\F(h,k,m))r(h,k,m)+\F(h,k,m)c(h), c(h) \right) &\text{if } k\leq K,\\
    c(h) &\text{if }k=K.
    \end{cases}
\end{align*}
Thus, we have $U_1^1(h,k)= U_1^2(h,k),~\forall h,k.$ As shown in the proof of \Cref{mono-hk-dis}, $V_1^1(h,k,m)$ and $V_1^2(h,k,m)$ are nonincreasing in $h$ and $k$. Hence, $U_1^1(h,k)$ and $U_1^2(h,k)$ are nonincreasing in $h$ and $k$ as well.

Since $\K_2\succeq_{st} \K_1$, by \Cref{puterman472-dis}, 
\begin{align*}
    &~~~~\sum_{k\in S_K}\left(   U_1^1(h,k)\K_1(k|h) -U_1^2(h,k)\K_2(k|h)\right)\\
    &=\sum_{k\in S_K}U_1^1(h,k)\left(\K_1(k|h) -\K_2(k|h)\right)\geq 0 ,~\forall h\in S_H,
\end{align*}
i.e., $v^1_1(h)\geq v^2_1(h),~\forall h \in S_H$.

\textbf{Induction step:} now assume that $v^1_n(h)\geq v^2_n(h)$ and $V_n^1(h,k,m)\geq V_n^2(h,k,m)$ for all $h,k,$ and $m$. We want to show $V_{n+1}^1(h,k,m)\geq V_{n+1}^2(h,k,m)$ for all $h,k,$ and $m$. By inspecting \Cref{value-iteration-dis}, we note that it suffices to show
\begin{align}
\begin{split}\label{2instance-1}
    \sum_{h'\in S_H}v^1_n(h')\H(h'|h) \geq \sum_{h'\in S_H}v^2_n(h')\H(h'|h),
\end{split}\\
\begin{split}\label{2instance-2}
    \sum_{h'\in S_H}v^1_n(h')\Q(h'|h) \geq \sum_{h'\in S_H}v^2_n(h')\Q(h'|h).
\end{split}
\end{align}

We will show \Cref{2instance-1}, and \Cref{2instance-2} can be proved in the same way. We write $v_n^i(h)=\sum_{k\in S_k} U_n^i(h,k) \\ \times \K_i(k|h),~i=1,2$. Since $V_n^1(h,k,m)\geq V_n^2(h,k,m)$ for all $h,k,$ and $m$, $U_n^1(h,k)\geq U_n^2(h,k)$ for all $h$ and $k$. Then,
\begin{align*}
    &~~~~\sum_{h'\in S_H}v^1_n(h')\H(h'|h) - \sum_{h'\in S_H}v^2_n(h')\H(h'|h)\\
    &=\sum_{h'\in S_H}\left(\sum_{k\in S_k} U_n^1(h',k) \K_1(k|h')\right)\H(h'|h) - \sum_{h'\in S_H}\left(\sum_{k\in S_k} U_n^2(h',k) \K_2(k|h')\right)\H(h'|h)\\
    &=\sum_{h'\in S_H}\H(h'|h)\sum_{k\in S_K} \left(  U_n^1(h',k)\K_1(k|h') -U_n^2(h',k)\K_2(k|h')\right)\\
    &\geq \sum_{h'\in S_H}\H(h'|h)\sum_{k\in S_K}\left(   U_n^2(h',k)\K_1(k|h') -U_n^2(h',k)\K_2(k|h')\right).
\end{align*}

As shown in the proof of \Cref{mono-hk-dis}, both $V_n^1(h,k,m)$ and $V_n^2(h,k,m)$ are nonincreasing in $k$ for any $n$. Thus, both $U_n^1(h,k)$ and $U_n^2(h,k)$ are nonincreasing in $k$. Since $\K_2\succeq_{st} \K_1$, by \Cref{puterman472-dis}, 
\begin{align*}
    \sum_{k\in S_K}\left(   U_n^2(h',k)\K_1(k|h') -U_n^2(h',k)\K_2(k|h')\right)\geq 0 ,~\forall h'\in S_H.
\end{align*}
Therefore, $\sum_{h'\in S_H}v^1_n(h')\H(h'|h) \geq \sum_{h'\in S_H}v^2_n(h')\H(h'|h)$.

It follows that $V_{n+1}^1(h,k,m)\geq V_{n+1}^2(h,k,m),~\forall h,k,$ and $m$. Then, $U_{n+1}^1(h,k)\geq U_{n+1}^2(h,k)$ for all $h$ and $k$, and
\begin{align*}
    v^1_{n+1}(h)-v^2_{n+1}(h)&= \sum_{k\in S_K}\left(U_{n+1}^1(h,k)\K_1(k|h) -U_{n+1}^2(h,k)\K_2(k|h)\right)\\
    &\geq \sum_{k\in S_K}\left(U_{n+1}^2(h,k)\K_1(k|h) -U_{n+1}^2(h,k)\K_2(k|h)\right)\\
    &\geq 0,~\forall h \in S_H,
\end{align*}
where the second inequality follows from \Cref{puterman472-dis} and $\K_2\succeq_{st} \K_1$. Therefore, $v^1_n(h)\geq v^2_n(h)$ and $V_n^1(h,k,m)\geq V_n^2(h,k,m),~\forall n$. Taking $n\rightarrow \infty$, we have $v^1(h)\geq v^2(h)$ and $V^1(h,k,m)\geq V^2(h,k,m),~\forall h,k,$ and $m$. $\blacksquare$

To prove \Cref{diff-trans}, we need the following lemma from \cite{alagoz2007determining}.
\begin{deflemma}\label{schaefer-lemma4}
Suppose transition probability functions $P(\cdot|\cdot),Q(\cdot|\cdot)$ on state space $S=\{1,\cdots,n\}$ satisfying $Q\succeq_{st} P$. Then the following inequalities hold for any function nonincreasing $f:\R\mapsto \R_{+}$: for $i,j=1,\cdots,n$,
\begin{enumerate}
    \item $\sum_{j\leq i}(P(j|i)-Q(j|i))f(j) \geq \sum_{j\leq i}(P(j|i)-Q(j|i))f(i)$.
    \item $\sum_{j> i}(P(j|i)-Q(j|i))f(j) \geq \sum_{j> i}(P(j|i)-Q(j|i))f(i+1)$.
\end{enumerate}
\end{deflemma}

\subsubsection*{Proof of \Cref{diff-trans}}
Prove by induction. Suppose that we solve $\Pi_1$ and $\Pi_2$ simultaneously using value iteration \Cref{value-iteration-dis} with $V_0^1(h,k,m)=V_0^2(h,k,m)=0,~\forall h,k,$ and $m$. Let $V^i_j$ be the value function of $\Pi_i,~i=1,2$ at the $j^{th}$ iteration.

\textbf{Initial step:} at the first iteration,
\begin{align*}
    V_1^1(h,k,m)&=V_1^2(h,k,m)\\
    &=\begin{cases}
    \max\left( (1-\F(h,k,m))r(h,k,m)+\F(h,k,m)c(h), c(h) \right) &\text{if } k\leq K,\\
    c(h) &\text{if }k=K.
    \end{cases}
\end{align*}
Thus, $v^1_1(h)=v^2_1(h), \forall h$.

\textbf{Induction step:} assume that $V_n^1(h,k,m)\geq V_n^2(h,k,m),~v_n^1(h)\geq v_n^2(h),~\forall h,k,$ and $m$. We want to show $V_{n+1}^1(h,k,m)\geq V_{n+1}^2(h,k,m),~v_{n+1}^1(h)\geq v_{n+1}^2(h),~\forall h,k,$ and $m$. By inspecting \Cref{value-iteration-dis}, we note that it suffices to show
\begin{align}
\begin{split}\label{2instance-1'}
    \sum_{h'\in S_H}v^1_n(h')\H_1(h'|h) \geq \sum_{h'\in S_H}v^2_n(h')\H_2(h'|h),
\end{split}\\
\begin{split}\label{2instance-2'}
    \sum_{h'\in S_H}v^1_n(h')\Q_1(h'|h) \geq \sum_{h'\in S_H}v^2_n(h')\Q_2(h'|h).
\end{split}
\end{align}

We will show \Cref{2instance-1'}, and \Cref{2instance-2'} can be proved in the same way. We have
\begin{align}\label{dom-arg}
\begin{split}
    &~~~~\sum_{h'\in S_H}v^1_n(h')\H_1(h'|h) - \sum_{h'\in S_H}v^2_n(h')\H_2(h'|h)\\
    &\geq \sum_{h'\leq h}v^2_n(h')\H_1(h'|h) +\sum_{h''> h}v^2_n(h'')\H_1(h''|h)\\ 
    &-\sum_{h'\leq h}v^2_n(h')\H_2(h'|h) -\sum_{h''> h}v^2_n(h'')\H_2(h''|h)\\
    &=\sum_{h'\leq h}v^2_n(h')\left(\H_1(h'|h)-\H_2(h'|h)\right) +\sum_{h''> h}v^2_n(h'')\left(\H_1(h''|h)-\H_2(h''|h)\right)\\
    &\geq v^2_n(h)\sum_{h'\leq h}\left(\H_1(h'|h)-\H_2(h'|h)\right) +v^2_n(h+1)\sum_{h''> h}\left(\H_1(h''|h)-\H_2(h''|h)\right)\\
    &= \left(v^2_n(h)-v^2_n(h+1)\right) \sum_{h'\leq h}\left(\H_1(h'|h)-\H_2(h'|h)\right)\\
    &\geq 0,
\end{split}
\end{align}
where the first inequality follows from the induction assumption that $v^1_n(h')\geq v^2_n(h'),~\forall h'$, the second inequality follows from \Cref{schaefer-lemma4} and $v_n^2$ being nonincreasing (shown in the proof of \Cref{mono-hk-dis}), and the second equality follows from the fact that
\begin{align*}
    \sum_{h'\leq h}\left(\H_1(h'|h)-\H_2(h'|h)\right) + \sum_{h''> h}\left(\H_1(h''|h)-\H_2(h''|h)\right)=0.
\end{align*}
The last equality holds because $v^2_n(h)\geq v^2_n(h+1)$ and $\sum_{h'\leq h}\left(\H_1(h'|h)-\H_2(h'|h)\right)\geq 0$, which follows from $\H_2\succeq_{st} \H_1$. Therefore, $V_{n+1}^1(h,k,m)\geq V_{n+1}^2(h,k,m),~\forall h,k,m$. Since $\Pi_1$ and $\Pi_2$ have the same $\K$ and $\M$, $v_{n+1}^1(h)\geq v_{n+1}^2(h)$. Taking $n\rightarrow \infty$, we have $v^1(h)\geq v^2(h)$ and $V^1(h,k,m)\geq V^2(h,k,m),~\forall h,k,$ and $m$. $\blacksquare$

\section{Selection of Parameters in Numerical Experiments}\label{appendb}
\setcounter{table}{0}
\renewcommand{\thetable}{B\arabic{table}}
\subsection{Definition of States and Transition Law}
The raw EPTS score is computed by the following formula \citep{unosEPTS}:
\begin{align*}
    \text{raw EPTS} &= (0.047-0.015\times \1\{\text{diabetes}\}) \times  (\text{age}-25)^+\\
    &+(0.398 -0.237\times \1\{\text{diabetes}\})\times  \text{number of prior organ transplantations}\\
    &+(0.315  -0.099\times \1\{\text{diabetes}\}) \times \log (\text{years on dialysis +1})\\
    &+(0.130-0.348\times \1\{\text{diabetes}\}) \times \1\{\text{no dialysis}\}\\
    &+1.262\times \1\{\text{diabetes}\}
\end{align*}
where $\1\{\cdot\}$ is the indicator function of an event and $x^+:=\max(x,0)$. Patients with lower EPTS scores are expected to have longer time of graft function from high-longevity kidneys, compared to patients with higher EPTS scores. The raw EPTS is converted to an EPTS score (ranging from $0$ to $100$) using the EPTS mapping table. An online calculator can be found at the OPTN website \citep{unosEPTScal}.


To define the probability of death, i.e., $\H(H+1|h)$, at each patient state, we find in the latest OPTN data report \citep{unos2022} that deaths per 100 patient years for kidney registrations during waiting is $6.91$ for patients over $65$, thus, the probability of death in a year is roughly $0.0691$, and therefore, we use $0.035$ as the probability of death for each epoch. We want to choose $\H(H+1|h),~\forall h$ to keep the arithmetic average of $\H(H+1|h)$ over $S_H$ 
close to $0.035$. Since older patients are more likely to die, we take $\H(H+1|h)$ to be increasing in $h$. 

If the patient chooses to wait and is alive, their state transition law is deterministic as shown in \Cref{EPTS-evolve} (\textbf{transition probability function} $\H$ can be defined accordingly). If the patient experiences a transplantation failure, we define the state transition law according to the change in their EPTS score, as shown in \Cref{EPTS-fail} (\textbf{transition probability function} $\Q$ can be defined accordingly). It is easy to check that \Cref{markovifr-dis,diffifr-dis,h-based-con1}, but not \Cref{h-based-con3}, are satisfied. However, \Cref{h-based-con3} would hold if we directly define patient state by EPTS score, as \Cref{EPTS-fail} indicates that there is a larger drop in EPTS score when patients with lower EPTS scores experience a transplant failure, compared with patients with higher EPTS scores.

\begin{table}
\centering
\tbl{Definition of patient state $h$ and corresponding EPTS scores.}{
\setlength\extrarowheight{1pt}
\begin{tabular}{|c|c|c|}
\hline
patient state $h$ & EPTS score \\ \hline
1         & 53     \\ \hline
2         & 60     \\ \hline
3         & 67     \\ \hline
4         & 72     \\ \hline
5         & 76     \\ \hline
6         & 80     \\ \hline
7         & 83     \\ \hline
8         & 86     \\ \hline
9         & 89    \\ \hline
10         & 91    \\ \hline
11         & 93    \\ \hline
12         & 94    \\ \hline
13         & 96    \\ \hline
14         & 97    \\ \hline
15         & 98    \\ \hline
16         & $99$ or greater   \\ \hline
17         & death    \\ \hline
\end{tabular}
}
\label{EPTS-evolve}
\end{table}


\begin{table}
\centering
\tbl{Patient state transition law after a transplantation failure.}{
\setlength\extrarowheight{1pt}
\begin{tabular}{|c|c|c|c|}
\hline
pre-transplant EPTS score & patient state $h$        & post-transplant EPTS score & patient state $h$ \\ \hline
53                        & 1               & 79                         & 6     \\ \hline
60                        & 2               & 85                         & 8     \\ \hline
67                        & 3               & 89                         & 9    \\ \hline
72                        & 4               & 92                         & 10    \\ \hline
76                        & 5               & 95                         & 12    \\ \hline
80                        & 6               & 97                         & 13    \\ \hline
83                        & 7               & 98                         & 14    \\ \hline
86 or greater             & 8 or greater    & 99 or greater              & 16    \\ \hline
\end{tabular}
}
\label{EPTS-fail}
\end{table}

We use the kidney donor profile index (KDPI) score to represent the \textbf{kidney state} $k$ and assume that kidney state $\{k_n\}_{n=1}^{\infty}$ form an i.i.d. sequence of random variables, independent of both patient state and mismatch level. KDPI, a scalar that combines ten donor factors including clinical parameters and demographics, is used in the kidney allocation system to measure the quality of deceased donor kidneys \citep{unosKDPI}. KDPI score ranges from $0$ to $100$. The average waiting time for a kidney offer is $2.13$ years according to the latest OPTN report \citep{unos2022}, so we assume that the patient's waiting time is a geometric random variable with mean $2.13$ years. Therefore, the probability that a patient gets an offer is $0.2347$ for each epoch (six months). We define four types of kidney state by their KDPI ranges. $\K$, \textbf{the distribution of the kidney state}, is obtained from \cite{unos2022} and shown in \Cref{KDPI-dis}. \Cref{mono-k-assump} holds, because we assume that the distribution of kidney state is independent of patient state.

\begin{table}
\centering
\tbl{The distribution of kidney state.}{
\setlength\extrarowheight{1pt}
\begin{tabular}{|c|c|c|}
\hline
KDPI range & kidney state $k$ & probability ($\%$) \\ \hline
0-20       & 1            & 4.91          \\ \hline
21-34      & 2            & 3.23         \\ \hline
35-85      & 3            & 12.06          \\ \hline
86-100     & 4            & 3.47          \\ \hline
Not available &5             & 76.53        \\ \hline
\end{tabular}
}
\label{KDPI-dis}
\end{table}

The \textbf{distribution of mismatch level} $\M$ in deceased donor kidney transplantation, also obtained from \cite{unos2022}, is given in \Cref{mismatch-dis}.

\begin{table}
\centering
\tbl{The distribution of mismatch level $\M$.}{
\setlength\extrarowheight{1pt}
\begin{tabular}{|c|c|c|}
\hline
mismatch level $m$  & probability ($\%$) \\ \hline
1                               & 4.92           \\ \hline
2                                & 1.04           \\ \hline
3                                & 1.92           \\ \hline
4                                & 14.37          \\ \hline
5                                & 28.06          \\ \hline
6                                & 32.54          \\ \hline
7                                & 14.14          \\ \hline
\end{tabular}
}
\label{mismatch-dis}
\end{table}

We assume in \Cref{sec4} that \textbf{the probability of a transplantation failure} depends only on $k$ and $m$, not on $h$, and we denote it by $\F(k,m)$. We use the six-month post-transplantation graft failure rate to represent $\F(k,m)$. For deceased donor kidney transplantation, the six-month post-transplantation graft survival rate is $97.1\%$ for perfect match, and $94.7\%$ for non-perfect match \citep{unos2022}. 

\begin{table}
\centering
\tbl{Six-month post-transplantation graft survival rate for zero and nonzero mismatch levels \citep{unos2022}.}{
\setlength\extrarowheight{1pt}
\begin{tabular}{|c|c|}
\hline
mismatch level $m$  & six month graft survival probability ($\%$) \\ \hline
$m=1$                  & 97.1           \\ \hline
$m>1$                & 94.7          \\ \hline
\end{tabular}
}
\label{mismatch-survival}
\end{table}

For deceased donor kidney transplantation, the six-month post-transplantation graft survival rate for different KDPI ranges is given in \Cref{KDPI-survival}.

\begin{table}
\centering
\tbl{Six-month post-transplantation graft survival rate for different KDPI ranges \citep{unos2022}.}{
\setlength\extrarowheight{1pt}
\begin{tabular}{|c|c|c|}
\hline
KDPI range & kidney state $k$ & graft survival probability ($\%$) \\ \hline
0-20       & 1            & 97.1           \\ \hline
21-34      & 2            & 95.1           \\ \hline
35-85      & 3            & 94.1           \\ \hline
86-100     & 4            & 91.6           \\ \hline
\end{tabular}
}
\label{KDPI-survival}
\end{table}

For each KDPI range, we assume that the six-month post-transplantation graft survival rate for $m=1$ and $m>1$ have the same ratio as \Cref{mismatch-survival}, and the six-month graft survival rate for each KDPI range in \Cref{KDPI-survival} is equal to the arithmetic average of the survival rate for $m=1$ and $m>1$. Then, \textbf{the probability of a transplantation failure} $\F(k,m)$ is given in \Cref{succ-trans}. We verify that \Cref{desenfail-dis} holds for $\F$ specified in \Cref{succ-trans}.

\begin{table}
\centering
\tbl{The probability ($\%$) of a transplantation failure $\F(k,m)$.}{
\setlength\extrarowheight{1pt}
\begin{tabular}{|c|c|c|}
\hline
$\F(k,m)$ & $m=1$ & $m>1$ \\ \hline
$k=1$       & 1.7            & 4.1          \\ \hline
$k=2$      & 3.7            & 6.1          \\ \hline
$k=3$      & 4.7            & 7.1          \\ \hline
$k=4$     & 7.3           & 9.5         \\ \hline
\end{tabular}
}
\label{succ-trans}
\end{table}

\subsection{Rewards}
We set the \textbf{intermediate reward} to be $0.5$ years, i.e., $c(h)=0.5,~\forall h$. We use the expected post-transplantation survival time to represent the \textbf{post-transplantation reward} $r(h,k,m)$. We assume that the patient post-transplantation survival time is a Poisson random variable (with unit to be a year). If we know the five-year post-transplantation patient survival rate, we can approximate the expected post-transplantation survival time with the mean of the corresponding Poisson random variable. For each EPTS-KDPI pair, the five-year post-transplantation patient survival rate can be found in \citep{bae2019can}. For each EPTS-KDPI range in \Cref{EPTS-KDPI-survival}, the five-year post-transplantation survival rate is approximated by the arithmetic average of survival rate at endpoints of the EPTS range with KDPI score to be the median of the KDPI range. 

\begin{table}
\centering
\tbl{The five-year post-transplantation patient survival rate ($\%$).}{
\setlength\extrarowheight{1pt}
\begin{tabular}{|c|c|c|c|c|}
\hline
\backslashbox{EPTS}{KDPI}  & 0-20  & 21-34 & 35-85 & 86-100 \\ \hline
53-54           & 87.5  & 86.8  & 83.85 & 76.55  \\ \hline
59-60           & 86    & 85.3  & 82.2  & 74.65  \\ \hline
67-68           & 83.75 & 83.05 & 79.75 & 72     \\ \hline
71-72           & 82.4  & 81.75 & 78.35 & 70.5   \\ \hline
75-76           & 80.95 & 80.3  & 76.8  & 68.85  \\ \hline
79-80           & 79.3  & 78.6  & 75.05 & 67.05  \\ \hline
83-84           & 77.6  & 76.8  & 73.05 & 65.15  \\ \hline
85-86           & 76.65 & 75.85 & 71.05 & 64.05  \\ \hline
89-90           & 74.45 & 73.8  & 69.85 & 61.95  \\ \hline
91-92          & 73.65 & 72.7  & 68.7  & 60.8   \\ \hline
93-94          & 72.6  & 71.6  & 67.45 & 59.6   \\ \hline
95-96          & 71.5  & 70.4  & 66.15 & 58.4   \\ \hline
97             & 70.6  & 69.5  & 65.2  & 57.5   \\ \hline
98             & 70    & 68.8  & 64.5  & 56.8   \\ \hline
99 or greater  & 69.4  & 68.2  & 63.8  & 56.2   \\ \hline
\end{tabular}
}
\label{EPTS-KDPI-survival}
\end{table}


To compute the post-transplantation reward, we also need to take into consideration the effect of mismatch level. We use the relative risk \citep{opelz2007effect} to measure relative contribution of HLA to the five-year patient survival rate.

\begin{table}
\centering
\tbl{Relative risk for different HLA-mismatch levels.}{
\setlength\extrarowheight{1pt}
\begin{tabular}{|c|c|}
\hline
Mismatch level $m$  & relative risk \\ \hline
1                  & 0.9        \\ \hline
2                  & 1        \\ \hline
3                  & 1.1        \\ \hline
4                  & 1.2             \\ \hline
5                  & 1.3       \\ \hline
6                  & 1.4        \\ \hline
7                  & 1.6        \\ \hline
\end{tabular}
}
\label{rr-hla}
\end{table}

We set mismatch level $m=5$ to be the reference and use the relative risk measure shown in \Cref{rr-hla}. Given EPTS score, KDPI score, and mismatch level, the five-year patient survival rate is obtained by the corresponding probability in \Cref{EPTS-KDPI-survival} divided by the relative risk in \Cref{rr-hla}.


\Cref{EPTS-KDPI-survival-0} and \Cref{EPTS-KDPI-survival-6} show the \textbf{expected post-transplantation patient survival time} for $m=1$ and $m=7$, respectively. When the EPTS or KDPI score is low, the expected post-transplantation survival time is much longer for $m=1$, compared with $m=7$. When the EPTS or KDPI score is high, the effect of mismatch level is less obvious. \Cref{transplant-reward-dis,wait-reward-dis} hold with the reward functions we used.

\begin{table}
\centering
\tbl{Expected post-transplantation patient survival time for $m=1$.}{
\setlength\extrarowheight{1pt}
\begin{tabular}{|c|c|c|c|c|}
\hline
\backslashbox{EPTS}{KDPI}  & 0-20  & 21-34 & 35-85 & 86-100 \\ \hline
53-54           & 12    & 11  & 10 & 8.5  \\ \hline
59-60           & 11  & 11  & 9.5 & 8.2  \\ \hline
67-68           & 9.9    & 9.7  & 9  & 7.9  \\ \hline
71-72           & 9.6 & 9.4 & 8.8 & 7.7     \\ \hline
75-76           & 9.3  & 9.1 & 8.5 & 7.6   \\ \hline
79-80           & 8.9 & 8.8  & 8.3  & 7.4  \\ \hline
83-84           & 8.7  & 8.5  & 8 & 7.2  \\ \hline
85-86           & 8.5  & 8.4  & 7.8 & 7.1  \\ \hline
89-90           & 8.3 & 8.1 & 7.7 & 6.9  \\ \hline
91-92           & 8.1 & 8  & 7.5 & 6.8  \\ \hline
93-94          & 8 & 7.9  & 7.4  & 6.7   \\ \hline
95-96          & 7.8  & 7.7  & 7.3 & 6.6   \\ \hline
97          & 7.7  & 7.6  & 7.2 & 6.6   \\ \hline
98          & 7.7  & 7.6  & 7.1 & 6.5   \\ \hline
99 or greater             & 7.6  & 7.5  & 7.1  & 6.5   \\ \hline
\end{tabular}
}
\label{EPTS-KDPI-survival-0}
\end{table}

\begin{table}
\centering
\tbl{Expected post-transplantation patient survival time for $m=7$.}{
\setlength\extrarowheight{1pt}
\begin{tabular}{|c|c|c|c|c|}
\hline
\backslashbox{EPTS}{KDPI}
  & 0-20  & 21-34 & 35-85 & 86-100 \\ \hline
53-54           & 6    & 5.9  & 5.8 & 5.5  \\ \hline
59-60           & 5.9  & 5.9  & 5.8 & 55  \\ \hline
67-68           & 5.8    & 5.8  & 5.7  & 5.4  \\ \hline
71-72           & 5.8 & 5.7 & 5.6 & 5.3     \\ \hline
75-76           & 5.7 & 5.7 & 5.6 & 5.3     \\ \hline
79-80           & 5.6  & 5.6 & 5.5 & 5.2   \\ \hline
83-84           & 5.6 & 5.6  & 5.4  & 5.1  \\ \hline
85-86           & 5.5  & 5.5  & 5.3 & 5.1  \\ \hline
89-90           & 5.5  & 5.4  & 5.3 & 5  \\ \hline
91-92           & 5.4 & 5.4 & 5.3 & 5  \\ \hline
93-94           & 5.4 & 5.4  & 5.2 & 4.9  \\ \hline
95-96          & 5.4 & 5.3  & 5.2  & 4.9   \\ \hline
97          & 5.3  & 5.3  & 5.1 & 4.8   \\ \hline
98          & 5.3  & 5.2  & 5.1 & 4.8   \\ \hline
99 or greater          & 5.3  & 5.2  & 5.1 & 4.8   \\ \hline
\end{tabular}
}
\label{EPTS-KDPI-survival-6}
\end{table}

\end{document}